\documentstyle[12pt,preprint,eqsecnum,aps]{revtex}
\tighten
\begin{document}
\preprint{MKPH-T-96-4}
\draft
\title{Virtual Compton scattering off the nucleon at low energies}
\author{S.\ Scherer\thanks{Supported by Deutsche Forschungsgemeinschaft}}
\address{Institut f\"ur Kernphysik, Johannes Gutenberg-Universit\"at,
55099 Mainz, Germany}
\author{A.\ Yu.\ Korchin}
\address{National Scientific Center Kharkov Institute of Physics and
Technology, 310108 Kharkov,\\ Ukraine}
\author{J.\ H.\ Koch}
\address{National Institute for Nuclear Physics and High-Energy Physics,
1009 DB Amsterdam,\\ The Netherlands}
\maketitle

\begin{abstract}
   We investigate the low-energy behavior of the four-point Green's function
$\Gamma^{\mu\nu}$ describing virtual Compton scattering off the nucleon.
   Using Lorentz invariance, gauge invariance, and crossing symmetry,
we derive the leading terms of an expansion of the operator in the
four-momenta $q$ and $q'$ of the initial and final photon, respectively.
   The model-independent result is expressed in terms of the electromagnetic
form factors of the free nucleon, i.e., on-shell information which
one obtains from electron-nucleon scattering experiments.
   Model-dependent terms appear in the operator at $O(q_\alpha q'_\beta)$,
whereas the orders $O(q_\alpha q_\beta)$ and $O(q'_\alpha q'_\beta)$
are contained in the low-energy theorem for $\Gamma^{\mu\nu}$, i.e.,
no new parameters appear.
   We discuss the leading terms of the matrix element and comment on
the use of on-shell equivalent electromagnetic vertices in the
calculation of ``Born terms'' for virtual Compton scattering.
\end{abstract}

\section{Introduction}
   Low-energy theorems (LET) play an important role in studies of properties
of particles.
   Based on a few general principles, they determine the leading terms of the
low-energy amplitude for a given reaction in terms of global,
model-independent properties of the particles.
   Clearly, this provides a constraint for models or theories of hadron
structure: unless they violate these general principles they must reproduce
the predictions of the low-energy theorem.
   On the other hand, the low-energy theorems also provide useful constraints
for experiments.
   Experimental studies designed to investigate particle properties beyond the
global quantities and to distinguish between different models must be carried
out with sufficient accuracy at low energies to be sensitive to the
higher-order terms not predicted by the theorems.
   Another option is, of course, to go to an energy regime where the
low-energy theorems do not apply anymore and model-dependent terms in the
theoretical predictions are important.

   The best-known low-energy theorem for electromagnetic interactions
is the theorem for ``Compton scattering'' (CS) of real photons off a nucleon
\cite{Low1,GellMann,Kazes}.
   Based on the requirement of gauge invariance, Lorentz invariance, and
crossing symmetry, it specifies the terms in the low-energy scattering
amplitude up to and including terms linear in the photon momentum.
   The coefficients of this expansion are expressed in terms of global
properties of the nucleon: its mass, charge and magnetic moment.
   In experiments, one can make the photon momentum -- the kinematical
variable in which one expands -- small to ensure the convergence of the
expansion and to allow for a direct comparison with the data.
   By increasing the energy of the photon one will become sensitive
to terms that depend on details of the structure of the nucleon beyond
the global properties.
   Terms of second order in the frequency, which are not determined by
this theorem, can be parameterized in terms of two new structure
constants, the electric and magnetic polarizabilities of the
nucleon (see, for example, \cite{Lvov}).

   As in all studies with electromagnetic probes, the possibilities to
investigate the structure of the target are much greater if virtual photons
are used.
   A virtual photon allows one to vary the three-momentum and energy
transfer to the target independently.
   Therefore it has recently been proposed to also use ``virtual Compton
scattering'' (VCS) as a means to study the structure of the nucleon
\cite{Audit,Brand,dHose}.
   The proposed reaction is $p(e,e'p)\gamma$, i.e., in addition
to the scattered electron also the recoiling proton is detected to
completely determine the kinematics of the final state consisting of a real
photon and a proton.
   It is the purpose of this note to extend the standard low-energy theorem for
Compton scattering of real photons to the general case where one or both
photons are virtual.
   The latter would be the case, e.g., in the reaction
$e^- + p \rightarrow  e^- + (e^-+e^+) + p$.
   We will refer to both possibilities as ``VCS''.

   There are several different approaches to derive the LET for Compton
scattering of real photons.
   One was first used by Low \cite{Low1}.
   It made use of the fact that in terms of ``unitarity diagrams'' the
scattering amplitude is dominated by the single-nucleon intermediate state.
   In such unitarity diagrams -- not to be confused with Feynman diagrams --
all intermediate states are on their mass shell \cite{Roman}.
   Lorentz invariance and gauge invariance then allow a prediction for the
amplitude to first order in the frequency.
   Another approach \cite{GellMann,Kazes}, first used by Gell-Mann and
Goldberger \cite{GellMann}, relies on a completely covariant description
in terms of the basic building blocks of electromagnetic vertices
and nucleon propagators.
   They split the amplitude into two classes, $A$ and $B$: class $A$ consists
of one-particle-reducible contributions that can be built up from dressed
photon-nucleon vertices and dressed nucleon propagators.
   Class $B$ contains all one-particle-irreducible two-photon diagrams, where
the second photon couples into the dressed vertex of the first one.
   Application of two Ward-Takahashi identities, one relating the
photon-nucleon vertex operator to the nucleon propagator, the other
relating the irreducible two-photon vertex to the dressed one-photon
vertex, then lead to the same result for the leading powers of the
low-energy amplitude.
   One can use yet another technique introduced by Low \cite{Low2}
to describe bremsstrahlung processes.
   This technique relies on the observation that poles in the photon momentum
can only be due to photon emission from external nucleon lines of the
scattering amplitude.
   Low's method has, in a modified form, also been widely used in the
framework of PCAC \cite{Adler}.

   So far, there have been only a few investigations of the general
VCS matrix element, since most calculations were restricted to the
Compton scattering of real photons.
   In \cite{Berg} electron-proton bremsstrahlung was calculated in
first-order Born approximation.
   The photon scattering amplitude, for one photon virtual and the other one
real, was analyzed in terms of 12 invariant functions of three scalar
kinematical variables.
   In \cite{Perrottet} it was shown that the general VCS matrix element
-- virtual photon to virtual photon -- requires 18 invariant amplitudes
depending on four scalar variables.
   In \cite{Schaefer} the reaction $\gamma p \to p+e^+e^-$ was investigated.
   Sizeable effects on the dilepton spectrum from the timelike electromagnetic
form factors of the proton were found.
   Very recently the low-energy behavior of the VCS matrix element
was investigated \cite{Guichon}.
   Using Low's approach \cite{Low2} the leading terms in the outgoing photon
momentum were derived.
   It was shown that the virtual Compton scattering amplitude at low energies
involves 10 ``generalized polarizabilities'' which depend on the absolute
value of the three-momentum of the virtual photon.
   These new polarizabilities were estimated in a nonrelativistic quark model.
   In \cite{Vanderhaeghen} the VCS amplitude was calculated in the framework
of a phenomenological Lagrangian including baryon resonances in the s and
u channel as well as $\pi^0$ and $\sigma$ exchange in the t channel.
   A prediction for the $|\vec{q}|$ dependence of the electromagnetic
polarizabilities $\alpha$ and $\beta$ was made.

   The purpose of this work is to identify, in an analogous form to
the real CS case, those terms of VCS which are determined
on the basis of only gauge invariance, Lorentz invariance, crossing
symmetry, and the discrete symmetries.
   In the following, we will refer to such terms not fixed by this
LET for simplicity as ``model dependent''.
   By introducing additional constraints, such as chiral symmetry,
also statements about these terms become possible.
   This is, however, beyond the scope of the present work.
   In our study of low-energy virtual Compton scattering, below the
onset of pion production, we will mainly work on the operator level.
   This allows us to work without specifying a particular Lorentz frame or
a gauge.
   We combine the method of Gell-Mann and Goldberger \cite{GellMann}
with an effective Lagrangian approach.
   Class $A$ is obtained in the framework of a general effective Lagrangian
describing the interaction of a single nucleon with the electromagnetic
field \cite{Lvov}.
   In the specific representation we choose, this turns out to be a simple
covariant and gauge invariant ``modified Born term'' expression,
involving on-shell Dirac and Pauli nucleon form factors $F_1$ and $F_2$.
   The Ward-Takahashi identities then allow us to determine the
leading-order term of the unknown class $B$ contribution in an expansion
in both the initial and final photon momenta.
   Furthermore, with the help of crossing symmetry a definite prediction can
be made concerning the order at which one expects model-dependent terms.

   Our work is organized as follows.
   We start out in Sec.\ II by outlining the general structure of the
VCS Green's function in the framework of Gell-Mann and Goldberger.
   We state the ingredients for the derivation of the LET, namely, crossing
symmetry and gauge invariance.
   Section III derives the LET for virtual photon Compton scattering
and we discuss the leading terms of the matrix element for the reaction
$e^-+p\to e^-+p+\gamma$ in the center-of-mass frame.
   As the notion of ``Born terms'' is important for the LET, we comment
on this aspect in Sec.\ IV and point out ambiguities that arise in their
definition, both for real and virtual photons.
   Our results are summarized and put into perspective in Sec.\ V.

\section{Structure of the Virtual Compton Scattering Tensor and Gauge
Invariance}
   In this section we will define the Green's functions and the kinematical
variables relevant for the discussion of VCS off the proton.
   We will consider the constraints imposed by the fundamental requirements of
gauge invariance, Lorentz invariance and crossing symmetry.
   We do this in the framework of a manifestly covariant description,
incorporating gauge invariance in its strong version, namely, in the form of
the Ward-Takahashi identities \cite{Ward,Takahashi}.
   The approach is similar to that of \cite{Kazes} using, however,
a somewhat more modern formulation.

   The electromagnetic three-point and four-point Green's functions are
defined as
\begin{eqnarray}
\label{tpgf}
G^\mu_{\alpha\beta}(x,y,z)&=&<\!0|T(\Psi_\alpha(x)\bar{\Psi}_\beta(y)
J^\mu(z))|0\!>,\\
\label{fpgf}
G^{\mu\nu}_{\alpha\beta}(w,x,y,z)&=&<\!0|T(\Psi_\alpha (w)\bar{\Psi}_\beta
(x) J^\mu(y) J^\nu(z))|0\!>,
\end{eqnarray}
where $J^\mu$ is the electromagnetic current operator in units of the
elementary charge, $e>0$, $e^2/4\pi=1/137$, and where $\Psi$ denotes a
renormalized interpolating field of the proton;
   $T$ denotes the covariant time-ordered product \cite{Itzykson}.
   Electromagnetic current conservation, $\partial_\mu J^\mu=0$, and
the equal-time commutation relation of the charge density operator
with the proton field,
\begin{equation}
\label{crjpsi}
[J^0(x),\Psi(y)]\delta(x^0-y^0)=-\delta^4 (x-y)\Psi(y),
\end{equation}
are the basic ingredients for deriving Ward-Takahashi identities
\cite{Ward,Takahashi}.

   Using translation invariance, the momentum-space Green's functions
corresponding to Eqs.\ (\ref{tpgf}) and (\ref{fpgf}) are defined through
a Fourier transformation,
\begin{eqnarray}
\label{tpgfm}
(2\pi)^4 \delta^4(p_f-p_i-q) G^\mu_{\alpha\beta}(p_f,p_i)&=&
\int d^4x d^4y d^4z\, e^{i(p_f\cdot x-p_i\cdot y- q\cdot z)}
G^\mu_{\alpha\beta}(x,y,z),\\
\label{fpgfm}
(2\pi)^4 \delta^4(p_f+q'-p_i-q) G^{\mu\nu}_{\alpha\beta}(P,q,q')&=
&\int d^4w d^4x d^4y d^4z  e^{i(p_f\cdot w-p_i\cdot x-q\cdot y+ q'
\cdot z)} G^{\mu\nu}_{\alpha\beta}(w,x,y,z),\nonumber\\
\end{eqnarray}
where $p_i$ and $p_f$ refer to the four-momenta of the initial and final
proton lines, respectively, $P=p_i+p_f$, and where $q$ and $-q'$ denote the
momentum transferred by the currents $J^\mu$ and $J^\nu$, respectively.
   We note that $G^\mu_{\alpha\beta}$ depends on two independent four-momenta,
e.g., $p_i$ and $p_f$.
   In particular, it is not assumed that these momenta obey the mass-shell
condition $p_i^2=p_f^2=M^2$.
   Similarly, $G^{\mu\nu}_{\alpha\beta}$ depends on three four-momenta
which are completely independent as long as one considers the general
off-mass-shell case.
   This will prove to be an important ingredient below when analyzing the
general structure of the VCS tensor.

   Finally, the truncated three-point and four-point Green's functions
relevant for our discussion of VCS are obtained by multiplying the external
proton lines by the inverse of the corresponding full (renormalized)
propagators,
\begin{eqnarray}
\label{itpgf}
\Gamma^{\mu}(p_f,p_i)&=&[iS(p_f)]^{-1}G^\mu(p_f,p_i)[iS(p_i)]^{-1},\\
\label{ifpgf}
\Gamma^{\mu\nu}(P,q,q')&=&[iS(p_f)]^{-1}G^{\mu\nu}(P,q,q')[iS(p_i)]^{-1},
\end{eqnarray}
where, for convenience, from now on we omit spinor indices.
   Using the definitions above, it is straightforward to obtain the
Ward-Takahashi identities
\begin{eqnarray}
\label{wt1}
q_\mu \Gamma^\mu(p_f,p_i)&=&S^{-1}(p_f)-S^{-1}(p_i),\\
\label{wt2}
q_\mu \Gamma^{\mu\nu}(P,q,q')&=&i\left(
S^{-1}(p_f)S(p_f-q)\Gamma^\nu(p_f-q,p_i)
-\Gamma^\nu(p_f,p_i+q)S(p_i+q)S^{-1}(p_i)\right).
\end{eqnarray}

   Following Gell-Mann and Goldberger \cite{GellMann}, we divide the
contributions to $\Gamma^{\mu\nu}$ into two classes, $A$ and $B$,
$\Gamma^{\mu\nu}=\Gamma^{\mu\nu}_A+\Gamma^{\mu\nu}_B,$
where class $A$ consists of the s- and u-channel pole terms;
   class $B$ contains all the other contributions.
   We emphasize that this procedure does not restrict the generality of the
approach.
   The separation into the two classes is such that all terms which are
irregular for $q^\mu\rightarrow 0$ (or $q'^\mu\rightarrow 0$) are contained in
class $A$, whereas class $B$ is regular in this limit.
   Strictly speaking, one also assumes that there are no massless particles in
the theory which could make a low-energy expansion in terms of kinematical
variables impossible \cite{Low1};
   furthermore, the contribution due to t-channel exchanges, such as a
$\pi^0$, has not been considered.

   The contribution from class $A$, expressed in terms of the full
renormalized propagator and the irreducible electromagnetic vertices,
reads
\begin{equation}
\label{gammaa}
\Gamma^{\mu\nu}_A=\Gamma^\nu(p_f,p_f+q')iS(p_i+q)
\Gamma^\mu(p_i+q,p_i)+\Gamma^\mu(p_f,p_f-q)iS(p_i-q') \Gamma^\nu(p_i-q',p_i).
\end{equation}
   Note that $\Gamma^{\mu\nu}_A$ is symmetric under crossing,
$q\leftrightarrow -q'$ and $\mu\leftrightarrow\nu$, i.e.,
\begin{equation}
\label{csgammaa}
\Gamma_A^{\mu\nu}(P,q,q')=\Gamma_A^{\nu\mu}(P,-q',-q).
\end{equation}
   Since also the total $\Gamma^{\mu\nu}$ is crossing
symmetric, this must also be true for the contribution of
class $B$ separately \cite{GellMann}.
   Using the Ward-Takahashi identity, Eq.\ (\ref{wt1}), one obtains the
following constraint for class $A$ as imposed by gauge invariance:
\begin{eqnarray}
\label{cgigammaa}
q_\mu \Gamma^{\mu\nu}_A(P,q,q')&=&i\left(\Gamma^\nu (p_f,p_f+q')
-\Gamma^\nu(p_i-q',p_i)
\right.\nonumber\\&&\left.
+S^{-1}(p_f)S(p_i-q')\Gamma^\nu(p_i-q',p_i)
-\Gamma^\nu(p_f,p_f+q')S(p_i+q)S^{-1}(p_i)\right)\nonumber\\
&\equiv&f^\nu_A(P,q,q').
\end{eqnarray}
   Similarly, contraction of $\Gamma^{\mu\nu}_A$ with $q'_{\nu}$ results in
\begin{eqnarray}
\label{cgigammaa2}
q'_\nu \Gamma^{\mu\nu}_A(P,q,q')&=&-i\left(\Gamma^\mu (p_f,p_f-q)
-\Gamma^\mu(p_i+q,p_i)
\right.\nonumber\\&&\left.
+S^{-1}(p_f)S(p_i+q)\Gamma^\mu(p_i+q,p_i)
-\Gamma^\mu(p_f,p_f-q)S(p_i-q')S^{-1}(p_i)\right)\nonumber\\
&=&-f^\mu_A(P,-q',-q),
\end{eqnarray}
which is, of course, the same constraint which one obtains from
Eq.\ (\ref{cgigammaa}) using the crossing-symmetry property of
$\Gamma^{\mu \nu}_{A}$:
\begin{equation}
\label{cgigammaa3}
q'_\nu \Gamma^{\mu\nu}_A(P,q,q')=-(-q'_\nu) \Gamma^{\nu\mu}_A(P,-q',-q)
=-f_A^\mu(P,-q',-q).
\end{equation}
   Combining Eqs.\ (\ref{wt2}) and (\ref{cgigammaa}) generates the following
constraint for the contribution of class $B$:
\begin{equation}
\label{cgigammab}
q_\mu \Gamma^{\mu\nu}_B=q_\mu (\Gamma^{\mu\nu}-\Gamma^{\mu\nu}_A)=
i\left(\Gamma^\nu(p_i-q',p_i)-\Gamma^\nu(p_f,p_f+q')\right),
\end{equation}
relating it to the one-photon vertex \cite{Kazes}.
   Once again, the second gauge-invariance condition, obtained by
contracting with $q'_{\nu}$, is automatically satisfied due to
crossing symmetry.

   The $4\times 4$ matrix $\Gamma^{\mu\nu}_B$ of class $B$ is a function of the
three independent four-momenta $P$, $q$, and $q'$.
   It is important to realize that the twelve components of these momenta
are independent variables only for the complete off-shell case,
i.e., if one allows for arbitrary values of $p_i^2$ and $p_f^2$.
   This will be important when making use of the constraints imposed
by gauge invariance.
   Using Lorentz invariance, gauge invariance, crossing symmetry, parity and
time-reversal invariance, it was shown in \cite{Perrottet} that the general
$\Gamma^{\mu\nu}$ for VCS off a free nucleon with both photons virtual
consists of 18 independent operator structures.
   The functions associated with each operator depend on four Lorentz
scalars, e.g., $q^2,q'^2,\nu=P\cdot q=P\cdot q'$, and $t=(p_i-p_f)^2$.
   When allowing the external nucleon lines to be off their mass shell, one
will have an even more complicated structure \cite{Melnitchouk}.

   However, in our derivation of the low-energy behavior of the
electromagnetic four-point Green's function we will not require the full
structure as discussed in \cite{Perrottet}.
   At low energies we expand $\Gamma^{\mu\nu}_B$ in terms of the four-momenta
$q^\mu$ and $q'^\mu$,
\begin{equation}
\label{expgammab}
\Gamma^{\mu\nu}_B=a^{\mu\nu}(P)+b^{\mu\nu,\rho}(P)q_\rho
+c^{\mu\nu,\sigma}(P)q'_\sigma+\cdots,
\end{equation}
where the coefficients are $4\times 4$ matrices and can be expressed
in terms of the 16 independent Dirac matrices
$1,\gamma_5,\gamma^\mu,\gamma^\mu \gamma_5, \sigma^{\mu\nu}$.
   An expansion of the type of Eq.\ (\ref{expgammab}) is expected to work below
the lowest relevant particle-production threshold, in this case the
pion-production threshold;
   we refer the reader to, e.g., \cite{SKF} where a similar discussion for
the case of pion photoproduction can be found.

   So far we have considered general features of the operators entering
into the description of VCS.
   It is clearly one of the advantages of using a covariant description of
the type of Eq.\ (\ref{expgammab}) that it neither uses a particular Lorentz
system nor a specific gauge.
   When referring to powers of $q$ or $q'$, we mean the ones coming from
the Dirac structures and their associated functions.
   This is different when one works on the level of nucleon matrix elements
or the invariant amplitude, where also kinematical variables from the spinors
or normalization factors enter into the power counting.

   We conclude this section by noting that the above framework can easily
be applied to VCS off a spin-0 particle, such as the pion.
   In that case $\Gamma^{\mu\nu}$, of course, has no complicated spinor
structure.
   The building blocks for class $A$ are simply the corresponding
irreducible, renormalized electromagnetic vertex for a spinless particle and
the full, renormalized propagator $\Delta(p)$ \cite{Fearing}.

\section{Low-energy behavior of VCS}
   In a two-step reaction on a single nucleon, such as
$\gamma^* N\rightarrow \gamma^* N$, the intermediate nucleon lines in the s-
and u-channel pole diagrams of class $A$ are off mass shell, while the
external nucleons are on shell.
   The early, manifestly covariant derivations of the low-energy theorem for
Compton scattering \cite{GellMann,Kazes} took into account that the associated
half-off-shell electromagnetic vertex of the nucleon has a different and more
complicated structure than the free vertex.
   However, it was shown that the model- and representation-dependent
properties of an off-shell nucleon do not enter in the leading terms of the
full Compton scattering amplitude when the irreducible two-photon amplitude
of class $B$ is included consistently.
   This was explicitly shown by expanding Eq.\ (\ref{gammaa}) to first order in
$q^\mu$ and by constructing the leading-order term of $\Gamma^{\mu\nu}_B$
with the help of Eq.\ (\ref{cgigammab}) and crossing symmetry or,
equivalently, the second gauge-invariance constraint.
   The final result for the amplitude was given in the laboratory frame and
in Coulomb gauge.

   In order to obtain the LET for VCS, we will proceed on the operator level
and combine the method of \cite{GellMann,Kazes} with ideas of an
effective Lagrangian approach to Compton scattering \cite{Lvov}.
   Let us first recall that the electromagnetic three-point and four-point
Green's functions of Eqs.\ (\ref{tpgf}) and (\ref{fpgf}) depend on the choice
of the interpolating field $\Psi$ of the proton, i.e., are
``representation dependent''.
   Therefore, the truncated Green's functions in momentum space,
Eqs.\ (\ref{itpgf}) and (\ref{ifpgf}), are in general not
directly related to observables, except for $p_i^2=p_f^2=M^2$.
   Consequently, the separation into class $A$ and class $B$ is necessarily
representation dependent, since the total Green's
function, Eq.\ (\ref{itpgf}), as well as the individual building blocks of
class $A$ are representation dependent; this has of course no effect on the
final on-shell result, which is representation independent.
   This was explicitly shown in \cite{Scherer1} for the case
of real Compton scattering off the pion.
   On the other hand, for any given appropriate interpolating field satisfying
the equal-time commutation relation of Eq.\ (\ref{crjpsi}) the Ward-Takahashi
identities, Eqs.\ (\ref{wt1}) and (\ref{wt2}), hold, providing important
consistency relations between the different Green's functions.
   In the following we will make use of these relations for arbitrary
$P$, $q$, and $q'$, which, in particular, includes arbitrary $p_i^2$
and $p_f^2$.
   Only at the end, the observable on-shell case will be considered.

\subsection{Derivation of the Low-Energy Theorem}
   We will derive this theorem by using a convenient representation,
the ``canonical form'', of the most general and gauge invariant effective
Lagrangian.
   In the framework of effective Lagrangians, a canonical form is defined as
a representation with the minimal number of independent structures (see,
e.g., \cite{Scherer2}).
   For the class $A$ terms, we need the electromagnetic vertex and the
propagator of the nucleon.
   Below the pion-production threshold, the Lagrangian for a single proton,
interacting with an electromagnetic field, can be brought into the
canonical form \cite{Lvov}
\begin{eqnarray}
\label{leff1}
{\cal L}_{\gamma NN}&=&\bar{\Psi}(i D\hspace{-.6em}/-M)\Psi
-e\sum_{n=1}^{\infty} ((-\Box)^{n-1}\partial^\nu F_{\mu\nu}) F_{1n}
\bar{\Psi}\gamma^\mu\Psi\nonumber\\
&&-\frac{e}{4M}\left(\kappa F_{\mu\nu}+\sum_{n=1}^\infty
((-\Box)^n F_{\mu\nu})F_{2n}\right)\bar{\Psi}\sigma^{\mu\nu}\Psi,
\end{eqnarray}
where $D_\mu\Psi=(\partial_\mu+ieA_\mu)\Psi$,
$F_{\mu\nu}=\partial_\mu A_\nu -\partial_\nu A_\mu$.
   The electromagnetic structure of the proton is accounted for through the
Dirac and Pauli form factors, $F_1$ and $F_2$,
respectively, which are expanded according to
\begin{equation}
F_1(q^2)=1+\sum_{n=1}^{\infty} (q^2)^n F_{1n}, \quad
F_2(q^2)=\kappa +\sum_{n=1}^{\infty} (q^2)^n F_{2n}, \quad
\kappa =1.79.
\end{equation}

   To this representation of the Lagrangian belongs, of course, a particular
canonical Lagrangian of order $e^2$ that generates the class $B$ terms.
   However, as will be seen below, we do not need to know it in detail
for this derivation.
   Clearly, Eq.\ (\ref{leff1}) is invariant under the gauge
transformation $\Psi\to\exp(-ie\alpha(x))\Psi$, and
$A_\mu\to A_\mu+\partial_\mu\alpha$.
   In order to arrive at Eq.\ (\ref{leff1}), use has implicitly
been made of the method of field transformations (see, e.g.,
\cite{Scherer2,Chisholm,Kamefuchi,Arzt}).
   It should be stressed that all ingredients needed for Eq.\ (\ref{leff1})
are on-shell quantities that can be determined model independently
from electron-proton scattering.
   Any explicit off-shell dependence of the irreducible three-point Green's
function has been transformed away and will thus not show up
in the class $A$ contribution.
   Such transformations, however, generate concomittant irreducible class $B$
terms for the amplitude that must be treated consistently (see \cite{Lvov}
for details).

   Using standard Feynman rules, the irreducible vertex associated with the
effective Lagrangian of Eq.\ (\ref{leff1}) is found to be
\begin{equation}
\label{geff}
\Gamma^\mu_{eff}(p_f,p_i)=\Gamma^\mu_{eff}(p_f-p_i)
=\gamma^\mu F_1(q^2)+\frac{1-F_1(q^2)}{q^2}q^\mu
q\hspace{-.45em}/ +i\frac{\sigma^{\mu\nu}q_\nu}{2M} F_2(q^2),
\quad q=p_f-p_i.
\end{equation}
   Since it is an important ingredient in our derivation of the LET,
we emphasize that the vertex of Eq.\ (\ref{geff}) satisfies the
Ward-Takahashi identity of Eq.\ (\ref{wt1});
   the corresponding propagator in the representation that yields
Eq.\ (\ref{leff1}) is the Feynman propagator of a point proton.
   As a consequence of gauge invariance, Eq.\ (\ref{leff1}) automatically
generates a term in the vertex proportional to $q^\mu$.

   We note that the effective Lagrangian approach provides a natural
explanation for the vertex of Eq.\ (\ref{geff}) which has previously been
used by several authors as a simple means to restore gauge invariance in the
form of the Ward-Takahashi identity (see, for example, \cite{Gross}).
   However, it is not an independent building block for any amplitude, but
must be used together with the corresponding irreducible class $B$
terms for the reaction in question!

   In principle, we could now proceed to construct the most general effective
Lagrangian relevant to generate the corresponding class $B$ terms for VCS.
   This would allow us, together with a calculation of the pole terms involving
the vertex of Eq.\ (\ref{geff}), to determine the model-dependent terms of
$\Gamma^{\mu\nu}$ at low energies.
   However, at this point it is more straightforward to apply the results
of the last section.
   We obtain for $\Gamma^{\mu\nu}_A$ in the framework of Eqs.\ (\ref{leff1})
and (\ref{geff})
\begin{equation}
\label{gammaaeff}
\Gamma^{\mu\nu}_{A,eff}=\Gamma^\nu_{eff}(-q')iS_F(p_i+q)
\Gamma^\mu_{eff}(q)+\Gamma^\mu_{eff}(q)iS_F(p_i-q') \Gamma^\nu_{eff}(-q').
\end{equation}
   Making use of Eq.\ (\ref{cgigammab}) to obtain the gauge-invariance
constraint for class $B$ in our representation, we see that it has the simple
form\footnote{In the following we omit the subscript ``$eff$''.}
\begin{equation}
\label{qgammab}
q_\mu \Gamma^{\mu\nu}_B=0.
\end{equation}
   This is due to the fact that the vertex of Eq.\ (\ref{geff})
depends on the momentum transfer, only.
   Note that Eq.\ (\ref{qgammab}) is still an operator equation, valid for
arbitrary $P$, $q$, and $q'$.

   Class $A$ is defined to contain for any representation the pole terms of
$\Gamma^{\mu\nu}$ which are singular in the limit $q^\mu\rightarrow 0$;
   we will come back to this point in the next section.
   We thus can make for class $B$ the following ansatz for the $q$ dependence:
\begin{equation}
\label{agammab}
\Gamma^{\mu\nu}_B(P,q,q')=a^{\mu,\nu}+b^{\mu\rho,\nu}q_\rho+
c^{\mu\rho\sigma,\nu}q_\rho q_\sigma+\cdots.
\end{equation}
   The coefficients $a^{\mu,\nu},\cdots$ are $4\times 4$ matrices which
have to be constructed from the 16 Dirac matrices, the metric tensor
$g^{\alpha\beta}$, the completely antisymmetric Levi-Civita pseudo tensor
$\epsilon^{\alpha\beta\gamma\delta}$, and the remaining independent
variables $P^\alpha$, and $q'^\alpha$.
   The form of these coefficients will be constrained by Lorentz invariance,
gauge invariance, crossing symmetry and the discrete symmetries, $C$, $P$,
and $T$.
   Contracting this ansatz with $q_\mu$, the condition for the class $B$
operator, Eq.\ (\ref{qgammab}), becomes
\begin{equation}
\label{giagammab}
q_\mu \Gamma^{\mu\nu}_B=a^{\mu,\nu}q_\mu + b^{\mu\rho,\nu}q_\rho q_\mu
+\cdots=0.
\end{equation}
Multiple partial differentiation of Eq.\ (\ref{giagammab}) with respect to
$q_\alpha$ results in the following conditions for the coefficients,
\begin{equation}
\label{concoeff}
a^{\mu,\nu}=0,\quad b^{\mu\rho,\nu}+b^{\rho\mu,\nu}=0,\quad
\sum_{\stackrel{6\, perm.}{(\mu,\rho,\sigma)}} c^{\mu\rho\sigma,\nu}=0,
\quad \cdots.
\end{equation}
   Since current conservation is expected to hold for arbitrary $q$,
the technique of partial differentiation to obtain conditions for
the coefficients can easily be extended to obtain constraints
for higher-order coefficients.
   The simple constraints of Eq.\ (\ref{concoeff}) are based on the fact that
$P^\alpha$, $q^\alpha$, and $q'^\alpha$ are independent variables;
   an implicit dependence would make matters more complicated.

   From $a^{\mu,\nu}=0$ we can already conclude that the operator of class
$B$ contains no contributions which involve powers of $q'$ only, without
powers of $q$.
   Taking Eq.\ (\ref{concoeff}) into account, we now expand with respect to
$q'$,
\begin{equation}
\label{gammab2}
\Gamma^{\mu\nu}_B(P,q,q')=B^{\mu\rho,\nu}q_\rho
+B^{\mu\rho,\nu\alpha}q_\rho q'_\alpha +\cdots
+C^{\mu\rho\sigma,\nu}q_\rho q_\sigma
+C^{\mu\rho\sigma,\nu\alpha}q_\rho q_\sigma q'_\alpha +\cdots.
\end{equation}
   If we apply crossing symmetry to Eq.\ (\ref{gammab2}) we find, as expected,
that indeed all terms vanish that involve powers of $q'$ only.
   Thus we find for the leading term of the model-dependent class $B$,
\begin{equation}
\label{bconstraints}
\Gamma_B^{\mu\nu}(P,q,q')=B^{\mu\rho,\nu\alpha}(P)q_\rho q'_\alpha
+O(qqq',qq'q'), \quad
B^{\mu\rho,\nu\alpha}=-B^{\rho\mu,\nu\alpha}=B^{\nu\alpha,\mu\rho},
\end{equation}
where the conditions for $B^{\mu\rho,\nu\alpha}$ result from gauge
invariance and crossing symmetry, respectively.
   To be specific, after imposing the constraints of Eq.\ (\ref{bconstraints})
and of the discrete symmetries one obtains three possible structures
for $B^{\mu\rho,\nu\alpha}$ on the {\em operator level}:
\begin{eqnarray}
\label{bexplicit}
B^{\mu\rho,\nu\alpha}(P)&=&i(g^{\mu\nu}g^{\rho\alpha}-g^{\rho\nu}g^{\mu\alpha})
f_1(P^2)\nonumber\\
&&+i(g^{\mu\nu}P^\rho P^\alpha + g^{\rho\alpha} P^\mu P^\nu
-g^{\rho\nu} P^\mu P^\alpha - g^{\mu\alpha}P^\rho P^\nu)f_2(P^2)\nonumber\\
&&+\epsilon^{\mu\rho\nu\alpha} \gamma_5 f_3(P^2),\quad \epsilon_{0123}=1.
\end{eqnarray}

   In summary, we have shown that  on the {\em operator level}
\begin {itemize}
\item the terms of order $O(q_{\alpha}^{-1})$, $O(q'^{-1}_{\alpha})$,
$O(1)$, $O(q_\alpha)$ and $O(q'_\alpha)$ are contained in
$\Gamma^{\mu\nu}_{A,eff}$, Eq.\ (\ref{gammaaeff}). They are therefore
determined model independently through on-shell quantities;
\item model-dependent terms first appear in the order $O(q_\alpha q'_\beta)$;
\item all operators which contain either only powers of $q$ or only powers
of $q'$ can entirely be obtained from this $\Gamma^{\mu\nu}_{A,eff}$.
\end {itemize}
   In the next section we will discuss the implications of these findings for
the on-shell VCS matrix element.

\subsection{Application}
   We now want to apply the above result to the observable case where the
nucleons are on mass shell, i.e., we consider the matrix element.
   For that purpose we first define the matrix element of $\Gamma^{\mu\nu}$
between positive-energy spinors as
\begin{equation}
\label{vmunu}
V^{\mu\nu}_{s_i s_f}(P,q,q')=\bar{u}(p_f,s_f)\Gamma^{\mu\nu}(P,q,q')u(p_i,s_i).
\end{equation}
   At this point we have to keep in mind that for the variables we use,
$P$, $q$, and $q'$, the on-shell condition $p_i^2=p_f^2=M^2$ is equivalent
to $P\cdot(q-q')=0$, and $P^2+(q-q')^2=4M^2$.
   In other words, the four-momenta chosen for the description of the
off-shell Green's function will no longer be independent for the on-shell
invariant amplitude.
   In particular, the distinction between powers of $q$ only or, respectively,
of $q'$ only in $\Gamma^{\mu\nu}$ is not valid anymore for the matrix element
since $q\cdot P=q'\cdot P$.

   Let us consider as an application the case where the initial photon is
virtual and spacelike and the final photon is real,
$\gamma^\ast(q,\epsilon)+p(p_i,s_i)\to\gamma(q',\epsilon')+p(p_f,s_f)$.
   The following discussion does not include the Bethe-Heitler
terms of the physical process $p(e,e'p)\gamma$, where the real photon is
radiated by the initial or final electron, since these terms are not part
of $\Gamma^{\mu\nu}$.
   For the final photon the Lorentz condition $q'\cdot\epsilon'=0$ is
automatically satisfied, and we are free to choose, in addition, the Coulomb
gauge $\epsilon'^\mu=(0,\vec{\epsilon}\,')$ which implies
$\vec{q}\,'\cdot\vec{\epsilon}\,'=0$.
   We write the invariant amplitude in the convention of
\cite{Bjorken} as ${\cal M}=-ie^2 \epsilon_\mu M^\mu$,
where $\epsilon_\mu=e\bar{u}\gamma_u u/q^2$ is the polarization
vector of the virtual photon.
With a suitable choice for the reference frame,
$q^\mu=(q_0,0,0,|\vec{q}|)$,
the Lorentz condition $q\cdot\epsilon=0$ and current conservation,
$q_\mu M^\mu=0$, may be used to reexpress the invariant amplitude as
\cite{Amaldi}
\begin{equation}
\label{invm}
{\cal M}=ie^2\left(\vec{\epsilon}_T\cdot\vec{M}_T
+\frac{q^2}{q_0^2}\epsilon_z M_z\right).
\end{equation}
   Note that as $q_0\to 0$, both $\epsilon_z$ and $M_z$ tend to zero such
that $\cal M$ in Eq.\ (\ref{invm}) remains finite.
   Making use that we now know the general structure of the first
undetermined operator, Eq.\ (\ref{bexplicit}), we can now consider
the matrix element up to second order in $q$ or $q'$.
   This will enable us to explicitly show how far the amplitude
is determined and where the first model-dependent terms enter.
   After a reduction from Dirac spinors to Pauli spinors the transverse
and longitudinal parts of Eq.\ (\ref{invm}) may be expressed in terms
of 8 and 4 independent structures, respectively \cite{Lvov,Guichon,Scherer3}:
\begin{eqnarray}
\label{mt}
\vec{\epsilon}_T\cdot\vec{M}_T&=&
\vec{\epsilon}\,'^\ast\cdot \vec{\epsilon}_T A_1
+i \vec{\sigma}\cdot(\vec{\epsilon}\,'^\ast\times\vec{\epsilon}_T) A_2
+(\hat{q}'\times\vec{\epsilon}\,'^\ast)\cdot(\hat{q}\times\vec{\epsilon}_T) A_3
+i \vec{\sigma}\cdot (\hat{q}'\times\vec{\epsilon}\,'^\ast)
\times(\hat{q}\times\vec{\epsilon}_T)A_4\nonumber\\
&&+i\hat{q}\cdot\vec{\epsilon}\,'^\ast
\vec{\sigma}\cdot(\hat{q}\times\vec{\epsilon}_T) A_5
+i\hat{q}'\cdot\vec{\epsilon}_T
\vec{\sigma}\cdot(\hat{q}'\times\vec{\epsilon}\,'^\ast) A_6
+i\hat{q}\cdot\vec{\epsilon}\,'^\ast
\vec{\sigma}\cdot(\hat{q}'\times\vec{\epsilon}_T) A_7\nonumber\\
&&+i\hat{q}'\cdot\vec{\epsilon}_T
\vec{\sigma}\cdot(\hat{q}\times\vec{\epsilon}\,'^\ast) A_8,\\
\label{ml}
\epsilon_z M_z&=&\epsilon_z \vec{\epsilon}\,'^\ast\cdot\hat{q} A_9
+i \epsilon_z\vec{\epsilon}\,'^\ast\cdot\hat{q}
\vec{\sigma}\cdot(\hat{q}'\times\hat{q})A_{10}
+i\epsilon_z\vec{\sigma}\cdot(\hat{q}\times\vec{\epsilon}\,'^\ast)A_{11}
+i\epsilon_z\vec{\sigma}\cdot(\hat{q}'\times\vec{\epsilon}\,'^\ast)A_{12}.
\nonumber\\
\end{eqnarray}
   The results for the functions $A_i$ in the CM system expanded up
to $O(2)$, i.e.\ $|\vec{q}\,'|^2$, $|\vec{q}\,'||\vec{q}|$ and $|\vec{q}|^2$,
are shown in Tables \ref{tablet} and \ref{tablel}.
   The derivation of the corresponding expressions is outlined in
Appendix A.

   In the amplitudes $A_5$ and $A_9$, terms of order $1/|\vec{q}\,'|$
appear.
   We have not further expanded them in $|\vec{q}|$ and kept, e.g., the
on-shell form factors $G_M$ and $G_E$.
   This was done mainly to stress that according to Low's theorem \cite{Low2}
these divergent terms are already entirely fixed.
   They are due to soft radiation off external lines, with the intermediate
line approaching the mass shell in the pole terms.
   They can be obtained from, e.g., the Born terms that contain the
same on-shell information (see however the caveats in Sec.\ IV).
   In order to uniquely identify these singular contributions we have expanded
all relevant expressions in terms of $|\vec{q}|$ and $|\vec{q}\,'|$.
   This is the reason why the argument of the form factors is
$Q^2=q^2|_{|\vec{q}\,'|=0}=-2M(E_i-M)$, where $E_i=\sqrt{M^2+\vec{q}^2}$.

   Not only the irregular contribution, but all terms in the amplitudes up
to terms linear in the photon momenta are uniquely determined through
well-known properties of the free nucleon: its mass, charge and magnetic
moment.
   These are the terms that make up the LET for VCS.
   Up to $O(2)$, the coefficients in addition also involve the electric
mean square radius, which we know from electron-proton scattering,
as well as the electric and magnetic polarizabilities which
also enter in real Compton scattering.
   The latter are the coefficients of the first model-dependent
terms in the expansion of the scattering matrix element.
   Their specific form, and thus the definitions of the polarizabilities,
depend on the particular representation one chooses for the
effective Lagrangian;
   the $O(2)$ results in the Tables are specific for the ``canonical form'',
which happens to be the standard choice for real Compton scattering.
   Note that to the order considered $A_7=-A_8=A_{10}$.
   The amplitudes $A_1$ and $A_9$ contain the electric polarizability
$\bar{\alpha}$ of the proton whereas $A_3$ involves the magnetic
polarizability $\bar{\beta}$ \cite{Lvov}.
   In terms of the functions $f_i(P^2)$ of Eq.\ (\ref{bexplicit}) they
are defined as
\begin{equation}
\label{abf}
\bar{\alpha}=e^2(f_1(4M^2)+4M^2f_2(4M^2)),\quad
\bar{\beta}=-e^2f_1(4M^2).
\end{equation}
   The function $f_3(4M^2)$ only contributes at order $O(3)$, since
$\gamma_5$ connects upper and lower component of positive energy
spinors which effectively leads to an additional power
of $|\vec{q}|$ or $|\vec{q}\,'|$ in the matrix element.

   In Table \ref{tabletr} we also show the results for the transverse
amplitudes in the CM frame for real Compton scattering.
   Since $A_5=-A_6$  and $A_7=-A_8$ the result effectively involves 6
independent amplitudes as required by time reversal invariance \cite{Lvov}.
   When comparing with other expressions in the literature
\cite{Low1,GellMann,Lvov,Scherer3}, one has to keep in mind
that they usually are given in the laboratory frame.
   This observation accounts, for example, for the difference of the
LET for $A_1$ and $A_3$ of real Compton scattering in the lab and the
CM frame.
   Note also that the $1/|\vec{q}\,'|$ singularity disappears in the real
Compton scattering limit, since $\omega=|\vec{q}|=|\vec{q}\,'|$ in this case.

   In conclusion, the low-energy behavior of the VCS matrix element for
$e^-+p\to e^-+p+\gamma$, expanded up to order $O(2)$ in $|\vec{q}|$
and $|\vec{q}\,'|$, contains, in addition to the structure coefficients
that enter into real Compton scattering, also the electric mean square radius
and the electric and magnetic Sachs form factors in the spacelike region
which all can be obtained from electron scattering off the proton.
   For the reaction $\gamma+p\to p+e^+ e^-$ \cite{Schaefer},
the analogous information for timelike momentum transfers is needed.
However, here one has to keep in mind that this information is not directly
accessible for $0<q^2<4M^2$.

\section{The Born terms}
   From the above discussion in a particular representation one might be
tempted to conclude that the leading terms for VCS -- on the operator level
or for the amplitude -- are in general simply given by ``Born terms''.
   However, some caveats are in order.
   First, one has to keep in mind that the low-energy behavior was obtained
from considerations involving the most general ansatz for the truncated
four-point Green's function.
   All terms one can think of are included into class $A$ and
class $B$.
   In, e.g., the derivation of \cite{Kazes} of the LET for Compton scattering,
where no special representation is chosen, it is clearly shown that {\it both}
class $A$ and class $B$ terms are needed.
   Implicit in all derivations is of course the assumption that a description
in terms of observable asymptotic hadronic degrees of freedom is sufficient
and complete at low energies.
   Even though subnucleonic degrees of freedom, quarks and gluons, are
ultimately the origin of the structure of the nucleon, it was shown
\cite{Leutwyler,DHoker} that an effective field-theory approach
\cite{Weinberg1,Gasser,GSS,Weinberg2} in terms of hadrons is
meaningful at low energies, thus allowing a classification into class
$A$ and class $B$.

   Secondly, there is an ambiguity concerning what exactly is meant by ``Born
terms'', once phenomenological form factors are introduced and the
result is not obtained from a microscopic Lagrangian.
   We will illustrate this ambiguity by considering different representations
of the photon-nucleon vertex.
   All representations contain the same information concerning the
electromagnetic structure of the on-shell nucleon, as obtained in
electron-nucleon scattering, but differ in the half-off-shell situation
encountered in the s- and u-channel pole terms of Compton scattering.
   This difference is, of course, accompanied by different class $B$ terms
such that the total result is the same.

   To explain the above points in more detail, we will first reconsider
the most general expression for class $A$ and, without going to
a special representation, identify those terms which contain the
irregular contribution for $q^\mu\rightarrow 0$ or $q'^\mu\rightarrow 0$.
   We find that these contributions can be expressed in terms
of on-shell quantities, in our case the Dirac and Pauli form factors $F_1$ and
$F_2$.
   We then show that the use of on-shell equivalent electromagnetic
vertices gives rise to the same results for the VCS matrix element as far
as the {\em irregular} terms are concerned, but not every choice will result
in ``Born terms'' which are gauge invariant.

\subsection{Irregular Contribution to the VCS matrix element}
   When calculating $V^{\mu\nu}_{s_i s_f}$, the irregular contribution
originates from the singularities of the propagators $S(p_i+q)$ and
$S(p_i-q')$ in the s and u channel, respectively.
   To be specific, below pion-production threshold the renormalized, full
propagator can be written as
\begin{equation}
\label{srsf}
S(p+q)=S_F(p+q)+ \mbox{regular terms},\quad (p+q)^2<(M+m_\pi)^2,
\end{equation}
where $S_F(p)$ denotes the free propagator of a nucleon with mass $M$.
   By ``regular terms'' we mean terms which have a well-defined, non-singular
limit for $q^\mu\rightarrow 0$.
   Thus, as long as we are only interested in the irregular terms, we can
simply replace the full, renormalized propagator by the free Feynman
propagator, $S_F$.

   The most general form of the irreducible, electromagnetic
vertex of the nucleon can be expressed in terms of 12 operators and
associated form functions \cite{Bincer,Naus}.
   These functions depend on three scalar variables, e.g., the squared
momentum transfer and the invariant masses of the initial and final nucleon
lines.
   A convenient parametrization of $\Gamma^{\mu}(p_f,p_i)$ is given by
\begin{equation}
\label{gmb}
\Gamma^{\mu}(p_f,p_i)=\sum_{\alpha,\beta=+,-}
\Lambda_\alpha(p_f)
\left(\gamma^\mu F^{\alpha\beta}_1
+i\frac{\sigma^{\mu\nu}q_\nu}{2M}F^{\alpha\beta}_2
+\frac{q^\mu}{M}F^{\alpha\beta}_3\right)
\Lambda_\beta(p_i),\,\,\,
F_i^{\alpha\beta}=F_i^{\alpha\beta}(q^2,p_f^2,p_i^2),
\end{equation}
where $q=p_f-p_i$, and $\Lambda_\pm(p)=(M\pm p\hspace{-.4em}/)/2M$.
   We have chosen a form which differs slightly from the convention of
\cite{Naus} in the definition of the projection operators and the
normalization of the $F_3$ form functions.
   We only need the following on-shell properties of the form functions:
\begin{equation}
\label{oslff}
F^{++}_1(q^2,M^2,M^2)=F_1(q^2),\quad
F^{++}_2(q^2,M^2,M^2)=F_2(q^2),\quad
F^{++}_3(q^2,M^2,M^2)=F_3(q^2)=0,
\end{equation}
where $F_1$ and $F_2$ are the standard Dirac and Pauli form factors and $F_3$
vanishes because of time-reversal invariance.
   One can also show that $F_3(q^2)$ vanishes due to current conservation.

   We now systematically isolate the irregular part of, e.g., the s-channel
pole diagram,\footnote{In the following we omit the indices $s_i$ and $s_f$.}
\begin{eqnarray*}
V^{\mu\nu}_{A,s}&=&\bar{u}(p_f)\Gamma^\nu(p_f,p_f+q')
iS(p_i+q)\Gamma^\mu(p_i+q,p_i)u(p_i)\\
&\approx&\bar{u}(p_f)\Gamma^\nu(p_f,p_f+q')
iS_F(p_i+q)\Gamma^\mu(p_i+q,p_i)u(p_i),
\end{eqnarray*}
where we made use of Eq.\ (\ref{srsf}), and where the symbol $\approx$ denotes
equality up to regular terms.
   We now insert Eq.\ (\ref{gmb}) for $\Gamma^\mu(p_i+q,p_i)$ and use
$\Lambda_-(p_i)u(p_i)=0$ and $\Lambda_+(p_i)u(p_i)=u(p_i)$ to obtain
\begin{eqnarray*}
V_{A,s}^{\mu\nu}\approx \bar{u}(p_f)\Gamma^\nu(p_f,p_f+q')iS_F(p_i+q)
&(&\Lambda_+(p_i+q)(\gamma^\mu F_1^{++}(q^2,s,M^2)+\cdots)\\
&+&\Lambda_-(p_i+q)(\gamma^\mu F_1^{-+}(q^2,s,M^2)+\cdots))u(p_i),
\end{eqnarray*}
where $s=(p_i+q)^2$.
   Since $S_F(p_i+q)\Lambda_-(p_i+q)$ results in a regular term, we have
\begin{displaymath}
V_{A,s}^{\mu\nu}\approx i\bar{u}(p_f)\Gamma^\nu(p_f,p_f+q')
\frac{p_i\hspace{-.7em}/\hspace{.3em}+q\hspace{-.45em}/+M}{s-M^2}
\Lambda_+(p_i+q)(\gamma^\mu F^{++}_1(q^2,s,M^2)+\cdots)u(p_i).
\end{displaymath}
   We now expand the form functions around $s_0=M^2$ and note that in the
higher-order terms the powers of $(s-M^2)$ cancel the denominator,
$s-M^2$, of the Feynman propagator and thus give rise to regular terms,
\begin{displaymath}
V_{A,s}^{\mu\nu}\approx \bar{u}(p_f)\Gamma^\nu(p_f,p_f+q')
iS_F(p_i+q)\Lambda_+(p_i+q)(\gamma^\mu F_1(q^2)+
i\frac{\sigma^{\mu\rho}q_\rho}{2M}F_2(q^2))u(p_i),
\end{displaymath}
where we made use of Eq.\ (\ref{oslff}).
   Using $\Lambda_+(p_i+q)=1-\Lambda_-(p_i+q)$ and then repeating the
same procedure for $\Gamma^\nu(p_f,p_f+q')$ we finally obtain
\begin{equation}
\label{tairr}
V_{A,s}^{\mu\nu}\approx \bar{u}(p_f,s_f)\Gamma^\nu_{F_1,F_2}(-q')
iS_F(p_i+q)\Gamma^\mu_{F_1,F_2}(q) u(p_i,s_i),
\end{equation}
where we introduced the abbreviation
\begin{equation}
\label{f1f2v}
\Gamma^\mu_{F_1,F_2}(q)=\gamma^\mu F_1(q^2)+i\frac{\sigma^{\mu\nu}
q_\nu}{2M}F_2(q^2).
\end{equation}
The procedure for the u-channel part, $V_{A,s}^{\mu\nu}$, is completely
analogous and we obtain for the sum of s- and u-channel contributions
\begin{eqnarray}
\label{map}
V_{A}^{\mu\nu}&\approx&\bar{u}(p_f) ( \Gamma^\nu_{F_1,F_2}(-q') iS_F(p_i+q)
\Gamma^\mu_{F_1,F_2}(q)
+\Gamma^\mu_{F_1,F_2}(q)iS_F(p_i-q') \Gamma^\nu_{F_1,F_2}(-q'))
u(p_i)\nonumber\\
&\equiv&V_{A'}^{\mu\nu},
\end{eqnarray}
which only involves on-shell quantities, the Dirac and Pauli form factors
and the nucleon mass.
   Explicit calculation, including the use of the Dirac equation, shows that
$V^{\mu\nu}_{A'}$ of Eq.\ (\ref{map}) is, in fact, identical with
evaluating Eq.\ (\ref{gammaaeff}) between on-shell spinors.
   With the help of either Eq.\ (\ref{cgigammaa}) or by straightforward
calculation it can easily be shown that Eq.\ (\ref{map}) is gauge invariant.
   This is a special feature when working with these particular vertex
operators and was essential for the derivation in \cite{Guichon}.
It is quite unexpected, since the electromagnetic vertex,
Eq.\ (\ref{f1f2v}), and the nucleon propagator in Eq.\ (\ref{map}) do not
satisfy the Ward-Takahashi identity (except at the real photon point,
$q^2 = 0)$:
\begin{equation}
\label{wtf1f2}
q_\mu \Gamma^\mu_{F_1,F_2}(p_f,p_i)=(p_f\hspace{-.9em}/\hspace{.3em}
-p_i\hspace{-.7em}/\hspace{.3em})F_1(q^2)\neq S_F^{-1}(p_f)
-S^{-1}_F(p_i).
\end{equation}

\subsection{Different representations of the on-shell vertex}
   We now turn to ``Born-term'' calculations involving other representations
of the nucleon current operator, i.e., other ways to introduce
the model-independent information about the electromagnetic
structure of the free nucleon into the calculation of the lowest-order terms.
   Two commonly used alternative ways to parametrize the nucleon current
operator are
\cite{Barnes}:
\begin{eqnarray}
\label{gegmv}
\Gamma^\mu_{G_E,G_M}(p_f,p_i)&=&\left(1-\frac{q^2}{4M^2}\right)^{-1}
\left(\frac{P^\mu}{2M}G_E(q^2)
+\frac{\gamma^\mu P\hspace{-.6em}/\hspace{.2em} q\hspace{-.45em}/
-q\hspace{-.45em}/ P\hspace{-.6em}/\gamma^\mu}{8M^2}G_M(q^2)\right),\\
\label{h1h2v}
\Gamma^\mu_{H_1,H_2}(p_f,p_i)&=&\gamma^\mu H_1(q^2)-\frac{P^\mu}{2M}
H_2(q^2),
\end{eqnarray}
where $P=p_i+p_f$ and
\begin{equation}
\label{relff}
G_E=F_1+\frac{q^2}{4M^2}F_2,\quad G_M=H_1=F_1+F_2,\quad
H_2=F_2.
\end{equation}
   Given Eqs.\ (\ref{relff}), it is straightforward to show the equivalence
for the free proton current, the matrix elements of
Eqs.\ (\ref{geff}), (\ref{f1f2v}), (\ref{gegmv}) and
(\ref{h1h2v}) between free positive-energy spinors.
   On the other hand, the current operators in Eqs.\ (\ref{f1f2v}),
(\ref{gegmv}) and (\ref{h1h2v}) do not satisfy the Ward-Takahashi identity
when used in conjunction with free propagators, not even at the real-photon
point (recall Eq.\ (\ref{wtf1f2})):
\begin{eqnarray}
\label{wtgegm}
q_\mu \Gamma^\mu_{G_E,G_M}(p_f,p_i)&=&\left(1-\frac{q^2}{4M^2}\right)^{-1}
\frac{p^2_f-p^2_i}{2M} G_E(q^2)
\neq S_F^{-1}(p_f)-S^{-1}_F(p_i),\\
\label{wth1h2}
q_\mu \Gamma^\mu_{H_1,H_2}(p_f,p_i)&=&(p_f\hspace{-.9em}/\hspace{.3em}
-p_i\hspace{-.7em}/\hspace{.3em})H_1(q^2)-\frac{p^2_f-p^2_i}{2M}H_2(q^2)
\neq S_F^{-1}(p_f)-S^{-1}_F(p_i).
\end{eqnarray}

   The ``Born terms'' calculated with these electromagnetic vertices
and free nucleon propagators are
\begin{eqnarray}
\label{btxy}
V^{\mu\nu}_{X}&=& \bar{u}(p_f,s_f) ( \Gamma^\nu_{X}(p_f,p_f+q')
iS_F(p_i+q) \Gamma^\mu_{X}(p_i+q,p_i)\nonumber\\
&&+\Gamma^\mu_{X}(p_f,p_f-q)iS_F(p_i-q') \Gamma^\nu_{X}(p_i-q',p_i))
u(p_i,s_i),
\end{eqnarray}
where $X$ denotes either the vertex of Eq.\ (\ref{gegmv}) in terms of
$G_E,G_M$ or the vertex of  Eq.\ (\ref{h1h2v}) involving $H_1,H_2$,
respectively.
   These ``Born terms'' are by construction crossing symmetric but in both
cases not gauge invariant,
\begin{eqnarray}
\label{gigegm}
q_\mu V^{\mu\nu}_{G_E,G_M}&=&i\left(1-\frac{q^2}{4M^2}\right)^{-1}
G_E(q^2)\bar{u}(p_f,s_f)(
\Gamma^\nu_{G_E,G_M}(p_f,p_i+q)-\Gamma^\nu_{G_E,G_M}(p_f-q,p_i)
\nonumber\\
&&\quad\quad\quad+ \Gamma^\nu_{G_E,G_M}(p_f,p_i+q)
\frac{q\hspace{-.45em}/}{2M}
+\frac{q\hspace{-.45em}/}{2M}
\Gamma^\nu_{G_E,G_M}(p_f-q,p_i))u(p_i,s_i),\\
\label{gih1h2}
q_\mu V^{\mu\nu}_{H_1,H_2}&=&
i (H_1(q^2)-H_2(q^2))\bar{u}(p_f,s_f)
(\Gamma^\nu_{H_1,H_2}(p_f,p_i+q)-\Gamma^\nu_{H_1,H_2}(p_f-q,p_i))u(p_i,s_i)
\nonumber\\
&-&i H_2(q^2)\bar{u}(p_f,s_f)
(\Gamma^\nu_{H_1,H_2}(p_f,p_i+q)\frac{q\hspace{-.45em}/}{2M}
+\frac{q\hspace{-.45em}/}{2M}\Gamma^\nu_{H_1,H_2}(p_f-q,p_f))u(p_i,s_i).
\end{eqnarray}
   With the help of Eqs.\ (\ref{gegmv}) and (\ref{h1h2v}) it is easy to see
that in each case the non-vanishing divergence is of order $q$.
   The fact that Eqs.\ (\ref{gigegm}) and (\ref{gih1h2}) do not contain
terms of the type $a\cdot q/b \cdot q$ suggests that in general the different
``Born terms'', based on equivalent parametrizations of the on-shell currents,
will differ by regular terms, only.
   It is of course conceivable that the amplitude could differ by irregular
terms, which are separately gauge invariant.
   Thus the above argument is not stringent and can only serve as a motivation
for the following claim which is essentially equivalent to Low's theorem
applied to the particular case of VCS:
   {\it Any ``Born-term'' calculation involving electromagnetic current
operators which correctly reproduce the on-shell electromagnetic current
of the nucleon will yield the same irregular contribution to the VCS matrix
element.}
   The key to the proof of this statement is the fact that any current operator
which transforms as a Lorentz four-vector can be brought into the form of
Eq.\ (\ref{gmb}).
   On-shell equivalence then amounts to the constraint that all operators
have the same on-shell limit of the $F^{++}_i$ form functions.
   In general, no statement can be made for either the other form functions
or off-shell kinematics.\footnote{Further conditions on the form functions
can be derived from the Ward-Takahashi identity and discrete symmetry
requirements \cite{Bincer,Naus}.}
   However, as we have seen above, the irregular contribution of class
$A$, and thus of the total VCS matrix element, only involves the on-shell
information contained in $F_1^{++}(q^2,M^2,M^2)$ and $F_2^{++}(q^2,M^2,M^2)$.
   Any information beyond this will give rise to regular terms.
   Thus the above statement is true.

   To illustrate the above, we bring for example the current operators
of Eq.\ (\ref{f1f2v}) and (\ref{h1h2v}), involving $F_1$ and $F_2$ or
$H_1$ and $H_2$, respectively, into the general form of Eq.\ (\ref{gmb}).
   By using $1=\Lambda_+(p)+\Lambda_-(p)$, the result for the commonly
used form of Eq.\ (\ref{f1f2v}) is given by
\begin{equation}
\label{mgf1f2v}
F_1^{\alpha\beta}(q^2,p_f^2,p_i^2)=F_1(q^2),\quad
F_2^{\alpha\beta}(q^2,p_f^2,p_i^2)=F_2(q^2),\quad
F_3^{\alpha\beta}(q^2,p_f^2,p_i^2)=0,\quad \alpha,\beta=+,-.
\end{equation}
   For the vertex given in Eq.\ (\ref{h1h2v}), we use
$\{\gamma^\mu,\gamma^\nu\}=2g^{\mu\nu}$
and momentum conservation at the vertex to rewrite
\begin{equation}
\frac{(p_i+p_f)^\mu}{2M}=\frac{p_f\hspace{-.9em}/\hspace{.3em}\gamma^\mu
+\gamma^\mu p_i\hspace{-.7em}/}{2M}-i\frac{\sigma^{\mu\nu}q_\nu}{2M}.
\end{equation}
   By inserting appropriate projection operators in the form
$p\hspace{-.4em}/= M(\Lambda_+(p)-\Lambda_-(p))$ and as above,
the vertex of Eq.\ (\ref{h1h2v}) can be expressed as
\begin{eqnarray}
\label{mgh1h2v}
\Gamma^{\mu}_{H_1,H_2}(p_f,p_i)&=&\Lambda_+(p_f)
\left(\gamma^\mu(H_1(q^2)-H_2(q^2))
+i\frac{\sigma^{\mu\nu}q_\nu}{2M}H_2(q^2)\right)
\Lambda_+(p_i)\nonumber\\
&+&\Lambda_-(p_f) \left(\gamma^\mu H_1(q^2)
+i\frac{\sigma^{\mu\nu}q_\nu}{2M}H_2(q^2)\right)\Lambda_+(p_i)\nonumber\\
&+&\Lambda_+(p_f)\left(\gamma^\mu H_1(q^2)
+i\frac{\sigma^{\mu\nu}q_\nu}{2M}H_2(q^2)\right)\Lambda_-(p_i)\nonumber\\
&+&\Lambda_-(p_f)\left(\gamma^\mu(H_1(q^2)+H_2(q^2))
+i\frac{\sigma^{\mu\nu}q_\nu}{2M}H_2(q^2)\right)
\Lambda_-(p_i).
\end{eqnarray}
   Of course, Eq.\ (\ref{mgh1h2v}) contains the same on-shell information
as Eq.\ (\ref{f1f2v}),  since the form factors satisfy Eq.\ (\ref{relff}).
   On the other hand, the expressions for all the other $F_i^{\alpha\beta}$
form functions differ.
   This is why the two ``Born term'' calculations based on these
two vertices differ with respect to regular terms.
   It is straightforward to extend the same considerations to the vertex
involving $G_E$ and $G_M$.

   In conclusion, we have shown that a calculation based on only
the ``Born terms'', built from any of the many possible on-shell
equivalent vertices and free nucleon propagators,
yields the same results for the {\em irregular} terms as the LET.
   Thus these ``Born terms''  will differ among each other through
regular terms.
   Furthermore, such ``Born terms'' are, in general, not gauge invariant;
an exception is the commonly used form involving the Dirac and Pauli form
factors $F_1$ and $F_2$.
   ``Generalized Born terms'' which are made gauge invariant by hand
through an ad hoc prescription also differ by regular terms.

   Important starting point for the derivation of the LET are
the irregular terms.
   It is thus also possible to split the total
VCS amplitude into ``Born terms'' plus ``rest'', instead of class $A$
and $B$ amplitudes, to arrive at the same result for the LET, i.e.\ up to
and including terms linear in the photon three-momenta.
   In general, this result will have contributions from ``Born terms'' and
the ``rest'' amplitude.
   If one uses a ``generalized Born amplitude'', all
the terms appearing in the LET are due to the expansion of the Born
amplitude.
   It is a well-known feature of soft-photon theorems that they cannot
make statements about terms which are separately gauge invariant
\cite{Ferrari,Bell,Fearing2}.
   One has to keep this in mind when discussing the structure-dependent
higher-order terms of VCS, i.e., one needs to specify which
Born or  class-$A$ terms have been separated.
   For example, in \cite{Guichon} the ``Born terms'' involving $F_1$ and
$F_2$ where separated since they provide without any further manipulation
a gauge-invariant amplitude.
   Then the residual part with respect to these particular ``Born terms''
was parametrized in terms of generalized polarizabilities.
   A natural question to ask is what would have happened had one separated a
different choice of ``generalized Born terms'' and defined generalized
polarizabilities in an analogous fashion with respect to the corresponding
residual amplitude.
   Obviously one would, in general, have found different numerical values for
the new generalized polarizabilities in order to obtain
the same total result.

\section{Conclusions}
   In studying the structure of composite strongly interacting
systems the electromagnetic interaction has been the traditional
and precise tool of investigation.
   In scattering of electrons from a nucleon, our knowledge is restricted
to two form factors that we can extract from experiments.
   Even though we have not yet been able to fully explain this information on
the basis of QCD, it is important to look for other observables allowing us to
test approximations to the exact QCD solution and effective, QCD inspired
models.
   Such effective models are expected to work especially at low energies.
   The electron accelerators now make it possible to study virtual Compton
scattering, which is clearly more powerful in probing the nucleon than the
scattering of real photons.
   In analyzing Compton scattering it is important to know how much
of the prediction is not a true test of a model, but fixed due to
general principles.
   These model-independent predictions for virtual Compton scattering were
the main topic of our discussion.

   The interest in virtual Compton scattering has also been due to another
aspect: When studying reactions on a nucleus, such as $(e,e'p)$, the
nucleon interacting with the electromagnetic probe is necessarily off
its mass shell.
   We have no model-independent information for the behavior of such a
nucleon and any conclusion about genuine medium modifications
must be based on firm theoretical ground how to deal with a single
nucleon under these kinematical circumstances.
   In fact, such a discussion depends very much on what one chooses
as an interpolating field for the intermediate, not observed nucleon.
    This clearly makes the ``off-shell behavior'' of the nucleon
representation dependent and unobservable.
    We discussed how certain features of the off-shell electromagnetic vertex
of the nucleon can be shifted into irreducible, reaction-specific
terms for the reaction amplitude.
   Two-step reactions on a free nucleon -- like (virtual) Compton
scattering  -- allow us to test many aspects of dealing with an
intermediate, off-shell nucleon under simpler circumstances, without
complications from, e.g., exchange currents or final state interactions.
   Understanding these aspects on the single nucleon level would seem a
prerequisite before any exotic claims can be made for nuclear reactions.

   We have studied the virtual Compton scattering first on the {\em operator}
level.
   Using the requirement of gauge invariance, as expressed by the
Ward-Takahashi identity, we derived constraints for the operator that
determine terms up to and including linear in the four-momenta
$q$ and $q'$.
   Also, we showed that on the operator level terms involving terms depending
only on $q$ or only on $q'$ are determined model independently in terms
of on-shell properties of the nucleon.

   To obtain these results, we used the method of Gell-Mann and Goldberger,
by splitting the contributions into general pole terms (class $A$) and
the one-particle irreducible two-photon contributions
(class $B$).
   We calculated class $A$ below pion-production threshold in
the framework of a specific representation for the most general effective
Lagrangian compatible with Lorentz invariance, gauge invariance and discrete
symmetries.
   This approach was introduced in \cite{Lvov} as a method for writing the
general structure of the Compton scattering amplitude
in a way that allows best to discuss its low-energy behavior.
   In this connection, we also showed the origin for a commonly used
form of the electromagetic vertex of the nucleon and stated the
consistency conditions for its use.

   After discussing the leading terms of the VCS operator, we considered the
{\em matrix element} for $\gamma^\ast p \to \gamma p$ in the photon-nucleon CM
frame.
   We found that the VCS amplitude up to and including terms linear
in the initial and final photon three-momentum can be expressed in terms
of information one can obtain from electron-proton scattering.
   This is the result analogous to the LET for the real Compton scattering
amplitude.
   As we also showed, the next order -- terms involving
$|\vec{q}\,'|^2$, $|\vec{q}\,'||\vec{q}|$, and $|\vec{q}|^2$ --
is also completely specified but now requires in addition also
the electromagnetic polarizabilities $\bar{\alpha}$ and $\bar{\beta}$
encountered in {\em real} Compton scattering.
   In other words, new structure-dependent information can only appear
at order three or higher in the three-momenta.
   Our results concern the expansion in terms of powers of both the
initial and final photon momentum.
   This allowed us to determine more terms than in \cite{Guichon},
where only the leading terms in the final momentum were concerned.
   On the other hand, by expanding in both momenta, the range
of applicability is smaller since {\em both} kinematical variables
should be small.

   We then considered different commonly used methods to include the
on-shell information contained in the electromagnetic form factors
in a ``Born-term'' calculation of the VCS matrix element.
   The fact that the ``Born terms'' calculated with $F_1$ and $F_2$ are
gauge invariant is not trivial, since the vertex and free propagator
do not satisfy the Ward-Takahashi identity.
   We explained why different on-shell equivalent forms for the
electromagnetic vertex operator lead to the same {\em irregular}
contribution in the VCS matrix element.
   We emphasized the importance of stating with respect to which
pole terms the structure-dependent terms are defined.

   Using only gauge invariance, Lorentz invariance, crossing
symmetry, and the discrete symmetries, we were able to make statements about
the low-energy behavior up to $O(2)$.
   Further conclusions can be reached by also taking into account
the constraints imposed by chiral symmetry.
   This would most naturally be done in the framework of Chiral
Perturbation Theory.
   In particular, predictions for the higher-order terms could be
obtained.

\section{Acknowledgements}
   The work of J.~H.~K.~is part of the research program of the Foundation for
Fundamental Research of Matter (FOM) and the National Organization for
Scientific Research (NWO).
   The collaboration between NIKHEF and the Institute for Nuclear Physics
at Mainz is supported in part by a grant from NATO.
   A.~Yu.~K.~thanks the Theory Group of NIKHEF-K for the kind hospitality and
the Netherlands Organization for International Cooperation in Higher
Education (NUFFIC) for financial support.
   A.~Yu.~K.~also would like to thank SFB 201 of the Deutsche
Forschungsgemeinschaft for its hospitality and financial support during
his stay at Mainz.
   S.~S.~would like to thank H.~W.~Fearing for useful discussions on the
soft-photon approximation.

\appendix
\section{}
   In this appendix we outline the calculation of the transverse and
longitudinal functions $A_i$ of Eq.\ (\ref{mt}) and (\ref{ml}),
respectively.
   For that purpose we split $M^\mu$ into its contributions from classes
$A$ and $B$, $M^\mu=M^\mu_A+M^\mu_B$.
   If we introduce
\begin{equation}
\label{a1}
F(-q')= {\epsilon}\hspace{-.4em}/'^{\ast}
(1+\frac{\kappa}{2M} q\hspace{-.45em}/')
\end{equation}
for the vertex involving the final real photon, the contribution of class
$A$ (see Eq.\ (\ref{gammaaeff})) reads:
\begin{equation}
\label{a2}
M^\mu_A=\bar{u}(p_f,s_f)\left(F(-q')S_F (p_i +q)
\Gamma^{\mu}_{eff}(q)+\Gamma^{\mu}_{eff}(q)S_F(p_i -q') F(-q')\right)
u(p_i,s_i),
\end{equation}
where $\Gamma^{\mu}_{eff}(q)$ is defined in Eq.\ (\ref{geff}).
   Applying the Dirac equation, $M^\mu_A$ can be written as
\begin{eqnarray}
\label{a3}
M^\mu_A&=&\bar{u}(p_f,s_f)\left(
2 \left( {p_f \cdot {{\epsilon}'}^{\ast}\over{s-M^2}}
+{p_i\cdot{\epsilon}'^{\ast}\over{u-M^2}} \right) \Gamma^\mu_{eff}(q)
+(1+\kappa)\left(
\frac{\epsilon\hspace{-.4em}/'^\ast q\hspace{-.45em}/'
\Gamma^\mu_{eff}(q)}{s-M^2}+
\frac{\Gamma^\mu_{eff}(q)
\epsilon\hspace{-.4em}/'^\ast q\hspace{-.45em}/'}{u-M^2}\right)\right.
\nonumber\\
&&\left. -\frac{\kappa}{M}\left(
\frac{p_f\cdot\epsilon '^\ast q\hspace{-.45em}/'\Gamma^\mu_{eff}(q)}{s-M^2}
-\frac{p_i\cdot\epsilon'^\ast \Gamma^\mu_{eff}(q) q\hspace{-.45em}/'}{u-M^2}
\right)
+\frac{\kappa}{2M}(\epsilon\hspace{-.4em}/'^\ast \Gamma^\mu_{eff}(q)
+\Gamma^\mu_{eff}(q)\epsilon\hspace{-.4em}/'^\ast)
\right) u(p_i,s_i),
\nonumber\\
\end{eqnarray}
where $s=(p_f +q')^2$ and $u=(p_i -q')^2$.
   Similarly, using Eq.\ (\ref{bconstraints}) and
(\ref{bexplicit}), $M^\mu_B$ can be written as
\begin{eqnarray}
\label{a4}
M^\mu_B&=&\bar{u}(p_f,s_f)\left(
(\epsilon'^{\ast\mu}q\cdot q'-q'^\mu\epsilon'^\ast\cdot q)f_1(P^2)\right.
\nonumber\\
&&+(\epsilon'^{\ast\mu} P\cdot q P\cdot q'
+P^\mu \epsilon'^\ast\cdot P q\cdot q'
-P^\mu\epsilon'^\ast\cdot q P\cdot q'
-q'^\mu \epsilon'^\ast\cdot P P\cdot q) f_2(P^2)\nonumber\\
&&\left.
-i\epsilon^{\mu\rho\nu\alpha}\epsilon'^\ast_\nu q_\rho q'_\alpha \gamma_5
f_3(P^2) +O(3)\right)u(p_i,s_i),
\end{eqnarray}
where $O(3)$ denotes terms of order 3 in $q$ or $q'$.

   The following considerations will be carried out in the center-of-mass
(CM) frame, where the four-momenta are given by
$q^\mu=(q_0,0,0,|\vec{q}|)$,
$p_i^\mu =(E_q ,-\vec{q})$, $q'^\mu=(|\vec{q}\,'|,\vec{q}\,')$, and
$p_f^\mu =(E_{q'} ,-\vec{q}\,')$.
   From energy conservation, $q_0+E_q=|\vec{q}\,'|+E_{q'}$, we infer
that we may choose $|\vec{q}|$, $|\vec{q}\,'|$, and $z=\hat{q}\cdot\hat{q}'$
as a set of independent variables.
   In terms of the CM variables the denominators of Eq.\ (\ref{a3}) are
proportional to $|\vec{q}\,'|$,
\begin{equation}
\label{a5}
s-M^2=2|\vec{q}\,'|(E_{q'}+|\vec{q}\,'|), \quad
u-M^2=-2|\vec{q}\,'|(E_q+z|\vec{q}|),
\end{equation}
and thus $M^\mu_A$ can be written as
\begin{equation}
\label{a6}
M^\mu_A={a^\mu(q,q')\over {|\vec{q}\,'|}} + b^\mu(q,q').
\end{equation}
   The functions $a^\mu(q,q')$ and  $b^\mu(q,q')$ are regular with
respect to $|\vec{q}\,'|$ and $|\vec{q}|$, and are given by
\begin{eqnarray}
\label{a7}
a^\mu(q,q')&=&K(\vec{q})\bar{u}(-\vec{q}\,',s_f )\Gamma^\mu_{eff}(q)
u(-\vec{q},s_i),\\
\label{a8}
b^\mu(q,q')&=& \bar{u}(-\vec{q}\,',s_f)\left(
(1+\kappa)
\left(
\frac{\epsilon\hspace{-.4em}/'^\ast n\hspace{-.5em}/'\Gamma^\mu_{eff}(q)}{
2(E_{q'}+|\vec{q}\,'|)}
-\frac{\Gamma^\mu_{eff}(q)\epsilon\hspace{-.4em}/'^\ast n\hspace{-.5em}/'}{
2(E_q+z|{\vec{q}|})}\right)\nonumber\right.\\
&&\left.-\kappa\frac{\vec{q}\cdot\vec{\epsilon}\,'^\ast}{M}
\frac{\Gamma^\mu_{eff}(q)n\hspace{-.5em}/'}{2(E_q+z|\vec{q}|)}
+\frac{\kappa}{2M}(\epsilon\hspace{-.4em}/'^\ast\Gamma^\mu_{eff}(q)
+\Gamma^\mu_{eff}(q)\epsilon\hspace{-.4em}/'^\ast)\right)
u(-\vec{q},s_i),
\end{eqnarray}
where we introduced $n'^\mu=(1,\hat{q}')$, and
\begin{equation}
\label{a9}
K(\vec{q})=-\frac{\vec{q}\cdot\vec{\epsilon}\,'^\ast}
{E_q +z|\vec{q}|}.
\end{equation}
   In order to obtain Eq.\ (\ref{a9}) we explicitly made use of the Coulomb
gauge for the final photon, namely, when using $p_f\cdot\epsilon'^\ast=0$
and $p_i\cdot\epsilon'^\ast=\vec{q}\cdot\vec{\epsilon}\,'^\ast$.

   We will now derive from Eqs.\ (\ref{a6}) - (\ref{a9}) the expansion of
$M^\mu_A$ in powers of $|\vec{q}|$ and $|\vec{q}\,'|$.
   According to Eq.\ (\ref{invm}) it is sufficient to treat the space
components of $M^\mu$.
   Since the structure coefficients appear at $O(q q')$ and higher in
the operator, we expand the contributions up to and including
$|\vec{q}\,'|^2$, $|\vec{q}\,'||\vec{q}|$ and $|\vec{q}|^2$.
   When expanding the electromagnetic vertex we make use of the following
relations
\begin{eqnarray}
\label{a10}
q^\mu&=&Q^\mu +|\vec{q}\,'| g^{\mu 0}+ \cdots,\nonumber\\
q\hspace{-.45em}/&=&Q\hspace{-.6em}/+|\vec{q}\,'|\gamma_0 +\cdots,\nonumber\\
q^2&=&Q^2 +2Q_0 |\vec{q}\,'|+\cdots,
\end{eqnarray}
where $Q^\mu=q^\mu|_{\vec{q}'=0}=(M-E_i,\vec{q})$.
Furthermore, for $\vec{a}$ we need the expansion of the form factors
around $Q^2$:
\begin{equation}
\label{a11}
F_{1,2}(q^2 )=F_{1,2}(Q^2 ) +2Q_0 |\vec{q}\,'| F_{1,2}'(Q^2 )+ \cdots,
\end{equation}
with the notation $F'(x)=dF/dx$.
   Using the definition of $\Gamma^\mu _{eff}(q)$ of Eq.\ (\ref{geff}),
it is straightforward to obtain
\begin{equation}
\label{a12}
\vec{\Gamma}_{eff}(q)=\vec{\Gamma}_{00}+\vec{\Gamma}_{10}|\vec{q}\,'|
+\vec{\Gamma}_{01}|\vec{q}| +\cdots,
\end{equation}
where
\begin{eqnarray}
\label{a13}
&&\vec{\Gamma}_{00}=\vec{\gamma},\quad
\vec{\Gamma}_{10}=-\frac{\kappa}{4M}{[\vec{\gamma},\gamma_0]},\quad
\vec{\Gamma}_{01}=\frac{\kappa}{4M}{[\vec{\gamma},\vec{\gamma}\cdot\hat{q}]},
\quad
\vec{\Gamma}_{20}=F'_1(0)\vec{\gamma}-\frac{\kappa}{8M^2}
{[\vec{\gamma},\gamma_0]},\nonumber\\
&&\vec{\Gamma}_{11}=-\hat{q}\gamma_0 F'_1(0),\quad
\vec{\Gamma}_{02}=(-\vec{\gamma}+\vec{\gamma}\cdot\hat{q}\hat{q})F'_1(0)
+\frac{\kappa}{8M^2}{[\vec{\gamma},\gamma_0]},\cdots .
\end{eqnarray}
   Since the dependence on the momenta $\vec{q}$ and $\vec{q}\,'$ is also
contained in the initial and final nucleon spinors, respectively, we
expand them to the required order:
\begin{equation}
\label{a14}
u(-\vec{q})=(1+\frac{\vec{\gamma}\cdot\vec{q}}{2M}+\frac{|\vec{q}|^2}{8M^2}
+\cdots)u(0),\quad
\bar{u}(-\vec{q}\,')=\bar{u}(0)(1+\frac{\vec{\gamma}\cdot\vec{q}\,'}{2M}
+\frac{|\vec{q}\,'|^2}{8M^2}+\cdots),
\end{equation}
where from now on we suppress the spin indices.
   Finally, the expansion of the energy denominators reads
\begin{equation}
\label{a15}
\frac{1}{E_{q'}+|\vec{q}\,'|}=\frac{1}{M}-\frac{|\vec{q}\,'|}{M^2}
+\frac{|\vec{q}\,'|^2}{2M^3} +\cdots,
\quad
\frac{1}{E_q+z|\vec{q}|}=\frac{1}{M}-\frac{z|\vec{q}|}{M^2}
+\frac{(2z^2-1)|\vec{q}|^2 }{2M^3}+\cdots.
\end{equation}

   Let us first consider $\vec{a}(q,q')$ which we expand according to
\begin{equation}
\label{a16}
\vec{a}(q,q')=\vec{a}_{00}(Q)+|\vec{q}\,'|\vec{a}_{10}+
|\vec{q}\,'|^2\vec{a}_{20}+|\vec{q}\,'||\vec{q}|\vec{a}_{11}
+|\vec{q}\,'|^3\vec{a}_{30}+|\vec{q}\,'|^2|\vec{q}|\vec{a}_{21}
+|\vec{q}\,'||\vec{q}|^2\vec{a}_{12}+\cdots.
\end{equation}
   Using the relations of Eqs.\ (\ref{a10}) -- (\ref{a15}) we obtain:
\begin{eqnarray}
\label{a17}
\vec{a}_{00}(Q)&=&K(\vec{q})\bar{u}(0)\vec{\Gamma}_{eff}(Q)u(-\vec{q}),
\\
\label{a18}
\vec{a}_{11}&=&-\frac{\hat{q}\cdot\vec{\epsilon}\,'^\ast}{M}
\bar{u}(0)\left( \frac{\vec{\gamma}\cdot\hat{q}\,'}{2M}
\vec{\Gamma}_{00} + \vec{\Gamma}_{10}\right)u(0),
\\
\label{a19}
\vec{a}_{21}&=&-\frac{\hat{q}\cdot\vec{\epsilon}\,'^\ast}{M}
\bar{u}(0)\left(\frac{1}{8M^2}\vec{\Gamma}_{00} +
\frac{\vec{\gamma}\cdot \hat{q}\,' }{2M}\vec{\Gamma}_{10} +
\vec{\Gamma}_{20}\right)u(0),
\\
\label{a20}
\vec{a}_{12}&=&-\frac{\hat{q}\cdot\vec{\epsilon}\,'^\ast}{M}
\bar{u}(0)\left(-\frac{z}{M}
(\frac{\vec{\gamma}\cdot\hat{q}\,'}{2M}\vec{\Gamma}_{00}+\vec{\Gamma}_{10})
+\frac{\vec{\gamma}\cdot\hat{q}\,'}{2M}\vec{\Gamma}_{01}\right.\nonumber\\
&&\left. +\vec{\Gamma}_{11}+ ( \frac{\vec{\gamma}\cdot\hat{q}\,'}{2M}
\vec{\Gamma}_{00}+\vec{\Gamma}_{10} )
\frac{\vec{\gamma}\cdot\hat{q}}{2M} \right) u(0),\\
\label{a21}
\vec{a}_{j0}&=&0,\,\, j=1,2,\cdots.
\end{eqnarray}
   The last equation follows from $K(0)=0$.
   The function $\vec{a}_{00}(Q)$ is completely determined in terms
of the electromagnetic form factors $F_{1,2}(Q^2)$ (or $G_{E,M}(Q^2)$).
   Note that in Eq.\ (\ref{a17}) we keep all powers in $|\vec{q}|$, since
it will be multiplied with the $1/|\vec{q}\,'|$ singularity and there are
no other terms which can generate such a singularity.

   In order to determine $\vec{b}(q,q')$, we first expand $\vec{b}(q,q')$
in an analogous fashion to Eq.\ (\ref{a16})
\begin{equation}
\label{a22}
\vec{b}(q,q')=\vec{b}_{00}+|\vec{q}\,'|\vec{b}_{10}+|\vec{q}|\vec{b}_{01}
+|\vec{q}\,'|^2 \vec{b}_{20}+|\vec{q}\,'||\vec{q}|\vec{b}_{11}
+|\vec{q}|^2\vec{b}_{02}+\cdots.
\end{equation}
   Using the building blocks of Eq.\ (\ref{a10}) - (\ref{a15}) we find
for the coefficients $\vec{b}_{ij}$:
\begin{eqnarray}
\label{a23}
\vec{b}_{00}&=&
\frac{1}{2M}\bar{u}(0)\left((1+\kappa){[\epsilon'^\ast\hspace{-1em}/
\hspace{.5em} n'\hspace{-.75em}/\hspace{.5em},\vec{\Gamma}_{00}]}
+\kappa\{\epsilon'^\ast\hspace{-1em}/\hspace{.5em},\vec{\Gamma}_{00}\}
\right)u(0)=\frac{1}{2M}\bar{u}(0) X u(0),\\
\label{a24}
\vec{b}_{10}&=&
\frac{1}{2M}\bar{u}(0)\left((1+\kappa)({[\epsilon'^\ast\hspace{-1em}/
\hspace{.5em} n'\hspace{-.75em}/\hspace{.5em},\vec{\Gamma}_{10}]}
-\frac{1}{M}\epsilon'^\ast\hspace{-1em}/\hspace{.5em}n'\hspace{-.75em}/
\hspace{.5em}\vec{\Gamma}_{00})
+\kappa\{\epsilon'^\ast\hspace{-1em}/\hspace{.5em},\vec{\Gamma}_{10}\}
+\frac{\vec{\gamma}\cdot\hat{q}'}{2M}X \right)u(0)\nonumber\\
&=&\frac{1}{2M}\bar{u}(0)
\left(Y+\frac{\vec{\gamma}\cdot\hat{q}'}{2M} X\right)u(0),\\
\label{a25}
\vec{b}_{01}&=&
\frac{1}{2M}\bar{u}(0)\left((1+\kappa)({[\epsilon'^\ast\hspace{-1em}/
\hspace{.5em} n'\hspace{-.75em}/\hspace{.5em},\vec{\Gamma}_{01}]}
+\frac{z}{M}\vec{\Gamma}_{00}
\epsilon'^\ast\hspace{-1em}/\hspace{.5em}n'\hspace{-.75em}/\hspace{.5em})
-\frac{\kappa}{M}\vec{\epsilon}\,'^\ast\cdot\hat{q}
\vec{\Gamma}_{00}n'\hspace{-.75em}/\hspace{.5em}
+\kappa\{\epsilon'^\ast\hspace{-1em}/\hspace{.5em},\vec{\Gamma}_{01}\}
+X\frac{\vec{\gamma}\cdot\hat{q}}{2M} \right)u(0)\nonumber\\
&=&\frac{1}{2M}\bar{u}(0)\left(Z+X \frac{\vec{\gamma}\cdot\hat{q}}{2M}
\right)u(0),\\
\label{a26}
\vec{b}_{20}&=&
\frac{1}{2M}\bar{u}(0)\left((1+\kappa)({[\epsilon'^\ast\hspace{-1em}/
\hspace{.5em} n'\hspace{-.75em}/\hspace{.5em},\vec{\Gamma}_{20}]}
-\frac{1}{M}\epsilon'^\ast\hspace{-1em}/\hspace{.5em}
n'\hspace{-.75em}/\hspace{.5em}\vec{\Gamma}_{10}
+\frac{1}{2M^2} \epsilon'^\ast\hspace{-1em}/\hspace{.5em}n'\hspace{-.75em}/
\hspace{.5em}\vec{\Gamma}_{00})
+\kappa\{\epsilon'^\ast\hspace{-1em}/\hspace{.5em},\vec{\Gamma}_{20}\}
\right.\nonumber\\
&&\left.+\frac{\vec{\gamma}\cdot\hat{q}'}{2M}Y+\frac{1}{8M^2}X\right)
u(0),\\
\label{a27}
\vec{b}_{11}&=&
\frac{1}{2M}\bar{u}(0)\left((1+\kappa)({[\epsilon'^\ast\hspace{-1em}/
\hspace{.5em} n'\hspace{-.75em}/\hspace{.5em},\vec{\Gamma}_{11}]}
-\frac{1}{M}\epsilon'^\ast\hspace{-1em}/\hspace{.5em}
n'\hspace{-.75em}/\hspace{.5em}\vec{\Gamma}_{01}
+\frac{z}{M}\vec{\Gamma}_{10}
\epsilon'^\ast\hspace{-1em}/\hspace{.5em}n'\hspace{-.75em}/\hspace{.5em})
-\frac{\kappa}{M}\vec{\epsilon}\,'^\ast\cdot\hat{q}\vec{\Gamma}_{10}
n'\hspace{-.75em}/\hspace{.5em}+\kappa
\{\epsilon'^\ast\hspace{-1em}/\hspace{.5em},\vec{\Gamma}_{11}\}
\right.\nonumber\\
&&\left.+Y\frac{\vec{\gamma}\cdot\hat{q}}{2M}
+\frac{\vec{\gamma}\cdot\hat{q}'}{2M}Z
+\frac{\vec{\gamma}\cdot\hat{q}'}{2M}X\frac{\vec{\gamma}\cdot\hat{q}}{2M}
\right)u(0),\\
\label{a28}
\vec{b}_{02}&=&
\frac{1}{2M}\bar{u}(0)\left(
(1+\kappa)({[\epsilon'^\ast\hspace{-1em}/
\hspace{.5em} n'\hspace{-.75em}/\hspace{.5em},\vec{\Gamma}_{02}]}
+\frac{z}{M}\vec{\Gamma}_{01}
\epsilon'^\ast\hspace{-1em}/\hspace{.5em}n'\hspace{-.75em}/\hspace{.5em}
+\frac{1-2z^2}{2M^2}\vec{\Gamma}_{00}
\epsilon'^\ast\hspace{-1em}/\hspace{.5em}
n'\hspace{-.75em}/\hspace{.5em})\right.\nonumber\\
&&\left.
-\frac{\kappa}{M}\vec{\epsilon}\,'^\ast\cdot\hat{q}(\vec{\Gamma}_{01}
-\frac{z}{M}\vec{\Gamma}_{00})n'\hspace{-.75em}/\hspace{.5em}
+\kappa\{\epsilon'^\ast\hspace{-1em}/\hspace{.5em},\vec{\Gamma}_{02}\}
+Z\frac{\vec{\gamma}\cdot\hat{q}}{2M}
+\frac{1}{8M^2}X\right)u(0).
\end{eqnarray}
   The reduction of the above expression to Pauli space is straightforward
but very tedious.
   The results are displayed in Tables \ref{tablea} and \ref{tableb}.

   Finally, it is straightforward to obtain the expansion of $\vec{M}_B$:
\begin{equation}
\label{a29}
\vec{M}_B=\bar{u}(0)\left(|\vec{q}\,'|^2\vec{\epsilon}\,'^\ast
\frac{\bar{\alpha}}{e^2}+|\vec{q}\,'||\vec{q}|(z\vec{\epsilon}\,'^\ast
-\hat{q}\cdot\vec{\epsilon}\,'^\ast\hat{q}')\frac{\bar{\beta}}{e^2}
+O(3)\right)u(0),
\end{equation}
where we have defined the (real) Compton polarizabilities as
\begin{equation}
\label{a30}
\bar{\alpha}=e^2(f_1(4M^2)+4M^2f_2(4M^2)),\quad
\bar{\beta}=-e^2f_1(4M^2).
\end{equation}
Due to the presence of the $\gamma_5$ matrix, the third function
$f_3(4M^2)$ only contributes at $O(3)$ at the level of the matrix
element.

\begin{table}
\caption{}{\label{tablet}
   Transverse functions $A_i$ of Eq.\ (\ref{mt}) in the CM frame.
   The functions are expanded in terms of $|\vec{q}\,'|$ and $|\vec{q}|$
of the final real and initial virtual photon, respectively.
$N_i=\sqrt{\frac{E_i+M}{2M}}$ is the normalization factor of the initial
spinor, where $E_i=\sqrt{M^2+|\vec{q}|^2}$.
   $G_E(q^2)=F_1(q^2)+\frac{q^2}{4M^2}F_2(q^2)$ and $G_M(q^2)=F_1(q^2)
+F_2(q^2)$ are the electric and magnetic Sachs form factors,
respectively.
   $r^2_E=6 G_E'(0)=(0.74\pm 0.02)\, \mbox{fm}^2$ is the electric mean square
radius \cite{Simon} and $\kappa=1.79$ the anomalous magnetic moment of the
proton.
   $Q^\mu$ is defined as $q^\mu|_{|\vec{q}\,'|=0}=(M-E_i,\vec{q})$,
$Q^2=-2M(E_i-M)$, and $z=\hat{q}\,'\cdot\hat{q}$.
$\bar{\alpha}$ and $\bar{\beta}$ are the electric and magnetic Compton
polarizabilities of the proton, respectively.}
\begin{tabular}{c|c}
$A_1$&
$-\frac{1}{M}+\frac{z}{M^2}|\vec{q}|
-\left(\frac{1}{8M^3}+\frac{r^2_E}{6M}-\frac{\kappa}{4M^3}
-\frac{\bar{\alpha}}{e^2}\right)
|\vec{q}\,'|^2
+\left(\frac{1}{8M^3}+\frac{r^2_E}{6M}-\frac{z^2}{M^3}+
\frac{(1+\kappa)\kappa}{4M^3}\right)|\vec{q}|^2$\\
\hline
$A_2$&
$\frac{1+2\kappa}{2M^2}|\vec{q}\,'|
-\frac{\kappa^2}{4M^3}|\vec{q}\,'|^2
+\frac{z\kappa}{2M^3}|\vec{q}\,'||\vec{q}|
-\frac{(1+\kappa)^2}{4M^3}|\vec{q}|^2$\\
\hline
$A_3$&
$-\frac{1}{M^2}|\vec{q}|
+\left(\frac{1}{4M^3}+\frac{\bar{\beta}}{e^2}\right)|\vec{q}\,'||\vec{q}|
+\frac{(3-2\kappa-\kappa^2)z}{4M^3}|\vec{q}|^2$\\
\hline
$A_4$&
$-\frac{(1+\kappa)^2}{2M^2}|\vec{q}|
-\frac{(2+\kappa)\kappa}{4M^3}|\vec{q}\,'||\vec{q}|
+\frac{(1+\kappa)^2 z}{4M^3}|\vec{q}|^2$\\
\hline
$A_5$&
$-\frac{N_i G_M(Q^2)}{(E_i+z|\vec{q}|)(E_i+M)}
\frac{|\vec{q}|^2}{|\vec{q}\,'|}
+\frac{(1+\kappa)\kappa}{4M^3}|\vec{q}|^2$\\
\hline
$A_6$&
$\frac{1+\kappa}{2M^2}|\vec{q}\,'|
-\frac{(1+\kappa)\kappa}{4M^3}|\vec{q}\,'|^2
-\frac{(1+\kappa)z}{2M^3}|\vec{q}\,'||\vec{q}|$\\
\hline
$A_7$&
$-\frac{1+3\kappa}{4M^3}|\vec{q}\,'||\vec{q}|$\\
\hline
$A_8$&
$\frac{1+3\kappa}{4M^3}|\vec{q}\,'||\vec{q}|$
\end{tabular}
\end{table}

\begin{table}
\caption{}{\label{tablel} Longitudinal functions $A_i$ of Eq.\ (\ref{ml})
in the CM frame. See caption of Table \ref{tablet}.}
\begin{tabular}{c|c}
$A_9$&
$\frac{N_i G_E(Q^2)}{(E_i+z|\vec{q}|)(E_i+M)}\frac{|\vec{q}|^2}{|\vec{q}\,'|}
-\frac{1}{M}
+\frac{z}{M^2}|\vec{q}|
-\left(\frac{1}{8M^3}+\frac{r^2_E}{6M}-\frac{\kappa}{4M^3}
-\frac{\bar{\alpha}}{e^2}\right)
|\vec{q}\,'|^2
+\left(\frac{1}{8M^3}+\frac{r^2_E}{6M}-\frac{z^2}{M^3}\right)
|\vec{q}|^2$\\
\hline
$A_{10}$&
$-\frac{1+3\kappa}{4M^3}|\vec{q}\,'||\vec{q}|$\\
\hline
$A_{11}$&
$-\frac{1+2\kappa}{2M^2}|\vec{q}\,'|
+\frac{\kappa^2}{4M^3}|\vec{q}\,'|^2
+\frac{(1+\kappa)z}{4M^3}|\vec{q}\,'||\vec{q}|
+\frac{1+2\kappa}{4M^3}|\vec{q}|^2$\\
\hline
$A_{12}$&
$\frac{(1+\kappa)z}{2M^2}|\vec{q}\,'|
-\frac{(1+\kappa)\kappa z}{4M^3}|\vec{q}\,'|^2
-\frac{(1+\kappa)(2z^2-1)}{4M^3}|\vec{q}\,'||\vec{q}|
-\frac{(1+\kappa)z}{4M^3}|\vec{q}|^2$
\end{tabular}
\end{table}

\begin{table}
\caption{\label{tabletr} Transverse functions $A_i$ in the CM frame
for both photons real: $q^2=q'^2=0$, $|\vec{q}|=|\vec{q}\,'|=\omega$.}
\begin{tabular}{c|c}
$A_1$&
$-\frac{1}{M}
+\frac{z}{M^2}\omega
+\left(-\frac{z^2}{M^3}+\frac{(2+\kappa)\kappa}{4M^3}
+\frac{\bar{\alpha}}{e^2}\right)\omega^2$\\
\hline
$A_2$&
$\frac{1+2\kappa}{2M^2}\omega
-\frac{1+2\kappa(1-z+\kappa)}{4M^3}\omega^2$\\
\hline
$A_3$&
$-\frac{1}{M^2}\omega
+\left(\frac{1+(3-\kappa(2+\kappa))z}{4M^3}+\frac{\bar{\beta}}{e^2}\right)
\omega^2$\\
\hline
$A_4$&
$-\frac{(1+\kappa)^2}{2M^2}\omega
+\frac{z+\kappa(2+\kappa)(z-1)}{4M^3}\omega^2$\\
\hline
$A_5$&
$-\frac{1+\kappa}{2M^2}\omega
+\frac{(2z+\kappa)(1+\kappa)}{4M^3}\omega^2$\\
\hline
$A_6$&
$\frac{1+\kappa}{2M^2}\omega
-\frac{(2z+\kappa)(1+\kappa)}{4M^3}\omega^2$\\
\hline
$A_7$&
$-\frac{1+3\kappa}{4M^3}\omega^2$\\
\hline
$A_8$&
$\frac{1+3\kappa}{4M^3}\omega^2$
\end{tabular}
\end{table}

\begin{table}
\caption{}{\label{tablea} Reduction of the coefficients $\vec{a}_{ij}$
of Eqs.\ (\ref{a17}) - (\ref{a20}) to Pauli space.
For the definition of the corresponding Pauli spin operators see
Eqs.\ (\ref{mt}) and (\ref{ml}). For further information see caption
of Table \ref{tablet}.}
\begin{tabular}{c|c|c|c|c}
&$\vec{a}_{00}(Q)/|\vec{q}\,'|$ &
$\vec{a}_{11}|\vec{q}|$ &
$\vec{a}_{21}|\vec{q}\,'||\vec{q}|$ &
$\vec{a}_{12}|\vec{q}|^2$\\
\hline
$A_1$&0&$\frac{z}{2M^2}|\vec{q}|$&
$-\frac{z\kappa}{4M^3}|\vec{q}\,'||\vec{q}|$&
$-\frac{z^2}{2M^3}|\vec{q}|^2$\\
\hline
$A_2$&0&0&0&0\\
\hline
$A_3$&0&$-\frac{1}{2M^2}|\vec{q}|$&
$\frac{\kappa}{4M^3}|\vec{q}\,'||\vec{q}|$&
$\frac{z}{2M^3}|\vec{q}|^2$\\
\hline
$A_4$&0&0&0&0\\
\hline
$A_5$&$-\frac{N_iG_M(Q^2)}{(E_i+z|\vec{q}|)(E_i+M)}
\frac{|\vec{q}|^2}{|\vec{q}\,'|}$&
0&0&$-\frac{\kappa}{4M^3}|\vec{q}|^2$\\
\hline
$A_6$&0&0&0&0\\
\hline
$A_7$&0&$\frac{1}{2M^2}|\vec{q}|$&$-\frac{\kappa}{4M^3}|\vec{q}\,'||\vec{q}|$&
$-\frac{z}{2M^3}|\vec{q}|^2$\\
\hline
$A_8$&0&0&0&0\\
\hline
$A_9$&$\frac{N_i G_E(Q^2)}{(E_i+z|\vec{q}|)(E_i+M)}
\frac{|\vec{q}|^2}{|\vec{q}\,'|}$&
$\frac{z}{2M^2}|\vec{q}|$&
$-\frac{z\kappa}{4M^3}|\vec{q}\,'||\vec{q}|$&
$-\frac{z^2}{2M^3}|\vec{q}|^2+\frac{r^2_E}{6M}|\vec{q}|^2$\\
\hline
$A_{10}$&0&$\frac{1}{2M^2}|\vec{q}|$&
$-\frac{\kappa}{4M^3}|\vec{q}\,'||\vec{q}|$&
$-\frac{z}{2M^3}|\vec{q}|^2$\\
\hline
$A_{11}$&0&0&0&0\\
\hline
$A_{12}$&0&0&0&0
\end{tabular}
\end{table}

\begin{table}
\caption{}{\label{tableb} Reduction of the coefficients $\vec{b}_{ij}$
of Eqs.\ (\ref{a23}) - (\ref{a28}) to Pauli space.
See captions of Tables \ref{tablet} and \ref{tablea}.}
\begin{tabular}{c|c|c|c|c|c|c}
&$\vec{b}_{00}$&$\vec{b}_{10}|\vec{q}\,'|$&$\vec{b}_{01}|\vec{q}|$&
$\vec{b}_{20}|\vec{q}\,'|^2$&$\vec{b}_{11}|\vec{q}\,'||\vec{q}|$&
$\vec{b}_{02}|\vec{q}|^2$\\
\hline
$A_1$&$-\frac{1}{M}$&0&
$\frac{z}{2M^2}|\vec{q}|$&
$-\left(\frac{1}{8M^3}+
\frac{r^2_E}{6M}-\frac{\kappa}{4M^3}
\right)|\vec{q}\,'|^2$&
$\frac{z\kappa}{4M^3}|\vec{q}\,'||\vec{q}|$&
$\left(\frac{1}{8M^3}+\frac{r^2_E}{6M}-\frac{z^2}{2M^3}\right.$\\
&&&&&&$\left.+\frac{\kappa (1+\kappa )}{4M^3}
\right)|\vec{q}|^2$\\
\hline
$A_2$&0&$\frac{1+2\kappa}{2M^2}|\vec{q}\,'|$&0&
$-\frac{\kappa^2}{4M^3}|\vec{q}\,'|^2$&
$\frac{z\kappa}{2M^3}|\vec{q}\,'||\vec{q}|$&
$-\frac{(1+\kappa)^2}{4M^3}|\vec{q}|^2$\\
\hline
$A_3$&0&0&$-\frac{1}{2M^2}|\vec{q}|$&0&
$\frac{1-\kappa}{4M^3}|\vec{q}\,'||\vec{q}|$&
$\frac{(1-2\kappa-\kappa^2)z}{4M^3}|\vec{q}|^2$\\
\hline
$A_4$&
0&
0&
$-\frac{(1+\kappa)^2}{2M^2}|\vec{q}|$&
0&
$-\frac{(2+\kappa)\kappa}{4M^3}|\vec{q}\,'||\vec{q}|$&
$\frac{(1+\kappa)^2z}{4M^3}|\vec{q}|^2$\\
\hline
$A_5$&
0&
0&
0&
0&
0&
$\frac{(2+\kappa)\kappa}{4M^3}|\vec{q}|^2$\\
\hline
$A_6$&
0&
$\frac{1+\kappa}{2M^2}|\vec{q}\,'|$&
0&
$-\frac{(1+\kappa)\kappa}{4M^3}|\vec{q}\,'|^2$&
$-\frac{(1+\kappa)z}{2M^3}|\vec{q}\,'||\vec{q}|$&
0\\
\hline
$A_7$&
0&
0&
$-\frac{1}{2M^2}|\vec{q}|$&
0&
$-\frac{1+2\kappa}{4M^3}|\vec{q}\,'||\vec{q}|$&
$\frac{z}{2M^3}|\vec{q}|^2$\\
\hline
$A_8$&
0&
0&
0&
0&
$\frac{1+3\kappa}{4M^3}|\vec{q}\,'||\vec{q}|$&
0\\
\hline
$A_9$&
$-\frac{1}{M}$&
0&
$\frac{z}{2M^2}|\vec{q}|$&
$-\left(\frac{1}{8M^3}+
\frac{r^2_E}{6M}-\frac{\kappa}{4M^3}
\right)|\vec{q}\,'|^2$&
$\frac{z\kappa}{4M^3}|\vec{q}\,'||\vec{q}|$&
$\left(\frac{1}{8M^3}-\frac{z^2}{2M^3}\right)|\vec{q}|^2$\\
\hline
$A_{10}$&
0&
0&
$-\frac{1}{2M^2}|\vec{q}|$&
0&
$-\frac{1+2\kappa}{4M^3}|\vec{q}\,'||\vec{q}|$&
$\frac{z}{2M^3}|\vec{q}|^2$\\
\hline
$A_{11}$&
0&
$-\frac{1+2\kappa}{2M^2}|\vec{q}\,'|$&
0&
$\frac{\kappa^2}{4M^3}|\vec{q}\,'|^2$&
$\frac{(1+\kappa)z}{4M^3}|\vec{q}\,'||\vec{q}|$&
$\frac{1+2\kappa}{4M^3}|\vec{q}|^2$\\
\hline
$A_{12}$&
0&
$\frac{(1+\kappa)z}{2M^2}|\vec{q}\,'|$&
0&
$-\frac{(1+\kappa)\kappa z}{4M^3}|\vec{q}\,'|^2$&
$-\frac{(1+\kappa)(2z^2-1)}{4M^3}|\vec{q}\,'||\vec{q}|$&
$-\frac{(1+\kappa)z}{4M^3}|\vec{q}|^2$
\end{tabular}
\end{table}
\end{document}